\documentclass[runningheads]{llncs}
\usepackage[utf8]{inputenc}
\usepackage{tabularx}
\newcolumntype{Y}{>{\centering\arraybackslash}X}
\usepackage{subcaption}
\usepackage{multirow}
\usepackage{graphicx}
\usepackage{xcolor, colortbl}
\usepackage{amsmath, mathrsfs}
\usepackage{amssymb}
\usepackage{paralist}
\usepackage{url}
\usepackage{enumitem}
\usepackage{fancyvrb}
\usepackage{algorithm}
\usepackage[backref=page]{hyperref}
\usepackage{algpseudocode}

\newcommand{\furl}[1]{\footnote{\scriptsize \url{#1}}}

\usepackage{forest}

\algnewcommand\algorithmicswitch{\textbf{switch}}
\algnewcommand\algorithmiccase{\textbf{case}}
\algnewcommand\algorithmicassert{\texttt{do}}
\algnewcommand\Assert[1]{\State \algorithmicassert(#1)}%
\algdef{SE}[SWITCH]{Switch}{EndSwitch}[1]{\algorithmicswitch\ #1\ \algorithmicdo}{\algorithmicend\ \algorithmicswitch}%
\algdef{SE}[CASE]{Case}{EndCase}[1]{\algorithmiccase\ #1}{\algorithmicend\ \algorithmiccase}%
\algtext*{EndSwitch}%
\algtext*{EndCase}%

\usepackage{pifont}
\newcommand{\xmark}{\ding{55}}%
\usepackage{multicol}
\usepackage{hhline}

\begin{document}
\title{\textbf{Dragoman}: Efficiently Evaluating Declarative Mapping Languages over Frameworks for Knowledge Graph Creation}
%
%
\author{Samaneh Jozashoori\inst{1,2}
Enrique Iglesias\inst{3}, 
Maria-Esther Vidal\inst{1,2,3}}
%
%
\institute{TIB Leibniz Information Center for Science and Technology, Germany
\\ 
\email{samaneh.jozashoori,maria.vidal@tib.eu}\\
\and
Leibniz University of Hannover \\
\and
L3S Research Center, Leibniz University of Hannover, Germany 
\\
\email{iglesias@l3s.de}
}
\maketitle         
%


\begin{abstract}
In recent years, there have been valuable efforts and contributions to make the process of RDF knowledge graph creation traceable and transparent; extending and applying declarative mapping languages is an example. One challenging step is the traceability of procedures that aim to overcome interoperability issues, a.k.a. data-level integration. In most pipelines, data integration is performed by ad-hoc programs, preventing traceability and reusability. However, formal frameworks provided by function-based declarative mapping languages such as FunUL and RML+FnO empower  expressiveness. Whether data processing or entity alignment, data-level integration can be defined as functions and integrated as part of the mappings performing schema-level integration. However, combining functions with the mappings introduces a new source of complexity that can considerably impact the required number of resources and execution time. We tackle the problem of efficiently executing mappings with functions and formalize the transformation of them into function-free mappings. These transformations are the basis of an optimization process that aims to perform an eager evaluation of function-based mapping rules. As a result, each function is executed once and efficiently reused. These techniques are implemented in a framework named Dragoman, providing, thus, the possibility to plan the optimized execution of functions and the materialization of reusable functions. We demonstrate the correctness of the transformations while ensuring that the function-free data integration processes are equivalent to the original one. The effectiveness of Dragoman is empirically evaluated in 230 testbeds composed of various types of functions integrated with mapping rules of different complexity. The outcomes suggest that evaluating function-free mapping rules reduces execution time in complex knowledge graph creation pipelines composed of large data sources and multiple types of mapping rules. The savings can be up to 75\%, suggesting that eagerly executing functions in mapping rules enables making these pipelines applicable and scalable in real-world settings.

\end{abstract}




\section{Introduction}
\label{sec:intro}

\noindent Knowledge graphs allow for representing data, metadata, and knowledge obtained from various sources in an integrated fashion using graphs. Knowledge graph creation can be defined as a data integration system (DIS) comprised of data sources, a unified schema, and mappings between the data sources and the concepts in the unified schema. Following the global view paradigm~\cite{lenzerini2002data}, mappings resolve the conflicts between different data sources schema, a.k.a. schema-level integration. The W3C standard mapping languages such as R2RML~\cite{das2012r2rml} and RML~\cite{dimou2014rml} allow for the declarative definition of mappings, assuring, thus, the transparency and the traceability of the process of knowledge graph creation. \\

\noindent In addition to the schema-level conflicts, data sources may have various levels of structuredness; they can suffer from various data quality issues or represent the same real-world entity differently. In general, knowledge graph creation pipelines need to include additional pre/post-processing blocks to resolve the interoperability issues at the data level, e.g., by using entity alignment. While isolated pre/post-processing blocks mostly lack traceability and reproducibility, integrating these blocks into the main pipeline of the knowledge graph creation forces traceability, which implies transparency. For this purpose, mapping languages have been extended to formalize data operations as functions that can be included either as programming scripts directly in the mapping rules~\cite{debruyne2016r2rml,junior2016funul,vu2019d} or as declarative representations (e.g., using the Function Ontology, FnO)~\cite{de2017declarative}. Albeit the advantages of integrating data operation functions in mappings, it can negatively affect the efficiency of the translation and interpretation of DIS to RDF. Hence, the data operation functions, as an element of data integration systems, need to be considered while optimizing and improving the efficiency of knowledge graph creation pipelines. Accordingly, this research is motivated by \textbf{i)} the necessity of having a transparent representation of data operations such as entity alignment, and \textbf{ii)} the importance of scaling up the process of knowledge graph creation to big data. \\

\begin{figure*}[t!]
\centering
\includegraphics[width=1.0\linewidth]{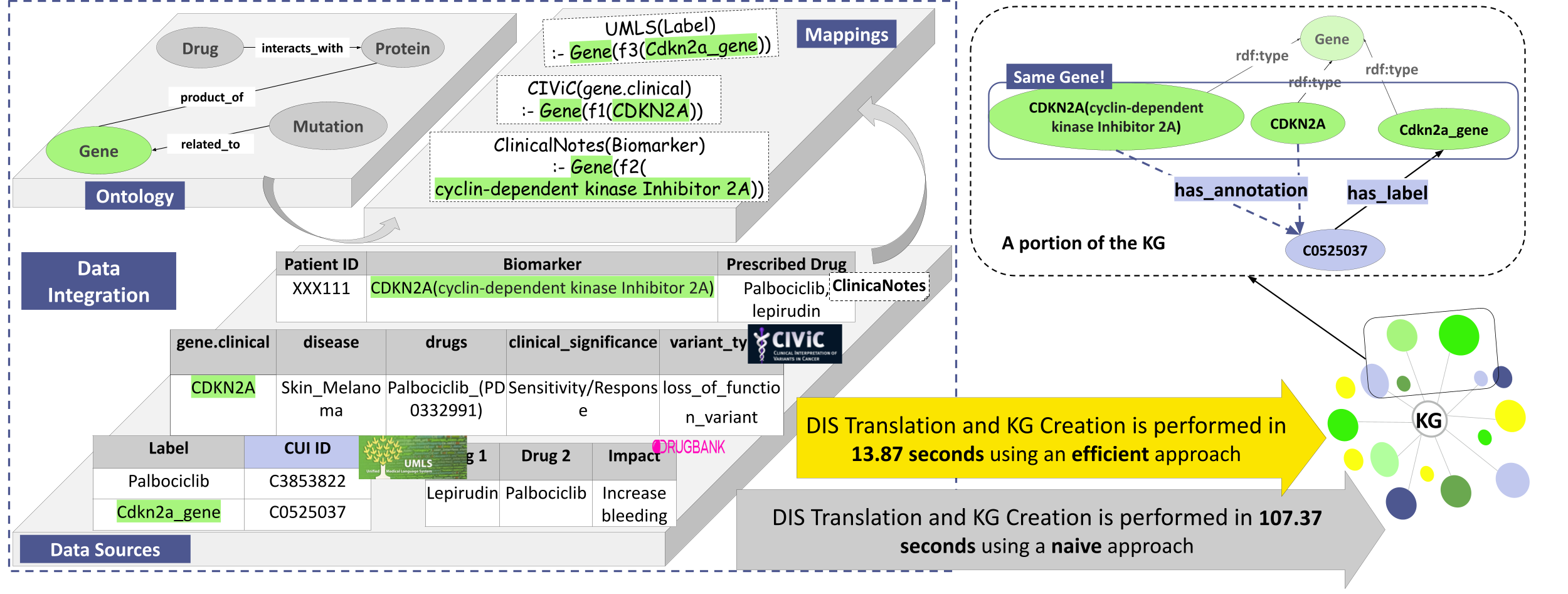}
\caption{\textbf{Motivating Example}. Three different representations of the same gene, CDKN2A, require to be aligned into the same entity while integrating the data from different sources and creating a knowledge graph. This entity alignment can be done using data operation functions to  annotate the data using control vocabularies. Applying a naive approach for executing the data operation functions and creating the knowledge graph can be expensive, while, an efficient approach can reduce the execution time of knowledge graph creation up to 75\% percent.}
\label{fig:motivation}
\end{figure*}

\noindent\textbf{Motivating Example.} 
The biomedical domain provides us with the basis to motivate our work. We aim to show the need 
of transparent representations of data operations executed during knowledge graph creation. CDKN2A is one of the critical loci of inactivation at both the germline and somatic mutations in patients with melanoma~\cite{tsao2012melanoma}. Many ongoing studies are established to investigate the mutations related to CDKN2A and its correlation to the diagnosis or prognosis of melanoma, prescribed treatments, interactions between different drugs, and the effectiveness of treatments in the presence of specific mutations. Therefore, a comprehensive insight of actionable knowledge can only be achieved by integrating and semantifying data from different studies residing in various sources. Nevertheless, it is essential to integrate mentioned data in a traceable and transparent manner; traceable, so that the observed results can be explained, and transparent, to verify the observed correlations or causation.\\

\noindent\autoref{fig:motivation} shows an example of integrating data derived from four different sources, i.e., CIViC\furl{https://civicdb.org/home}, DrugBank\furl{https://go.drugbank.com/}, UMLS~\cite{UMLS}, and clinical notes from hospitals into a knowledge graph. It can be observed that, in addition to the various attribute names used to represent gene data in different data sources, the representative values of the same gene, i.e., CDKN2A, also differ among the sources. Hence, mapping the gene values to the ``Gene'' concept in the ontology only provides the solution to the schema-level integration. The instances of the Gene class need to be aligned, performing entity alignment. For this purpose, one solution is to annotate the instances using standard vocabularies or metathesaurus concepts such as Concept Unique Identifier (CUI) from UMLS database~\cite{UMLS}; shown in purple in \autoref{fig:motivation}. The entity alignment task can be defined as a data operation function that accesses an engine performing Named Entity Recognition (NER) and Entity Linking and retrieves the annotation.
Owing to the existing declarative formalism of functions, entity alignment can be part of the main pipeline of the knowledge graph using functions in mappings. Nonetheless, a DIS translation as part of the knowledge graph creation - shown in \autoref{fig:motivation} - can be an expensive process in large heterogeneous data. Therefore, adding another layer, i.e., executing data operation functions, may increase the complexity. As shown in \autoref{fig:motivation}, following a naive approach, i.e., no specific optimization for function execution translating a DIS including entity alignment functions~\cite{jozashoori2022eablock} and 10k of data records, can be ten times more expensive (in terms of execution time) than an optimized approach. We face these data management challenges and propose an approach to scale up the materialized knowledge graph creation in the presence of data operation functions.\\

\noindent\textbf{Problem Statement.}
We address the problem of efficiently creating knowledge graphs from data integration systems (DIS), including data operation functions. Data processing and entity alignment can be defined as data operation functions and declared as part of the mappings in a declarative knowledge graph creation pipeline. Materializing a knowledge graph in the presence of large data sources is alone an expensive task. The incorporation of data operation functions as part of a DIS that defines a knowledge graph appends another level of complexity, which requires distinct optimization plans to scale up. Given a DIS, we aim to identify an equivalent function-free DIS that minimizes the execution time. The solution to this problem is to define a set of transformation rules.\\    

\noindent\textbf{Proposed Solution.} We propose a heuristic-based framework called Dragoman composed of a set of transformation rules to convert a given DIS involving data operation functions into a function-free one. Relying on an eager evaluation strategy, first, Dragoman recognizes all the data operation functions in mapping assertions and evaluates them. After the materialization of the functions, considering the mappings, Dragoman decides on the transformations required to transform an input DIS into a function-free one. These transformations are needed to ensure the correctness and completeness of the transformed DIS. In addition, Dragoman also performs transformations that result in the DIS that is both efficient and system-agnostic in terms of translation into RDF. In other words, any compliant engine can efficiently translate the transformed DIS into RDF triples. \\

\noindent\textbf{Contributions.} The contributions provided in this paper are the follows:
\begin{inparaenum}[\bf 1\upshape)]
\item Formalizing user-defined data operation functions as part of the DIS and declarative knowledge graph creation pipelines.
\item A set of transformation rules to convert a DIS with functions into a function-free DIS. 
\item Dragoman, a framework, that implements the proposed transformation rules and generates function-free RML-compliant knowledge graph creation pipelines. Following an eager evaluation strategy, Dragoman plans engine-agnostic execution plans of user-defined functions and RML mapping assertions.   
\item An exhaustive empirical evaluation of the performance of knowledge graph creation pipelines. In total, the outcomes of 230 testbeds are reported; they include various configurations that stress the process of knowledge graph creation~\cite{Chaves-FragaEIC19}.
\end{inparaenum}
\noindent This paper is organized into six additional sections.
Preliminaries, the formalization of user-defined data operation functions, and the semantics of mapping assertions are presented in section \ref{sec:preliminaries}. Section \ref{sec:approach} presents the problem of knowledge graph creation and discusses the proposed transformation rules and Dragoman, the proposed framework. Section \ref{sec:experiments} reports on the results of the empirical evaluation and related approaches are discussed in section \ref{sec:relatedwork}. Lastly, section \ref{sec:conclusions} summarizes lessons learned and outlines future directions.

\section{Mapping Assertions to Declaratively Specify Knowledge Graph Creation Pipelines}
\label{sec:preliminaries}
\noindent A logic program describes the world with a finite set of facts, a set of assertions about pieces in the world, and a set of rules allowing one to deduce facts from the known facts~\cite{ceri1989you}. Facts and rules can be represented as \textbf{Horn clauses}. Horn clauses with exactly one positive literal are \emph{definite clauses} and can be of general shape, $G_0 :- G_1,...,G_n$ where the left-hand side of the clause is called its body and the right-hand side is the head; if the body holds, then the head holds: $body(\overline{X}):-head(\overline{Y})$. $body(\overline{X})$ is a conjunction of predicates defined over terms, and $head(\overline{Y})$ is one predicate also defined over a set of \emph{terms}. We define the \emph{term} inductively as follows.

\noindent \textbf{Base Case.} i) Let $c$ be a constant. $c$ is a term. ii) let $X$ be a variable and $X$ be a term.

\noindent \textbf{Inductive Case.} Let $h$ be a functional symbol of arity $n$ and $t_1,...,t_n$ be \emph{term}s, then $h(t_1,...,t_n)$ is a \emph{term}.

\subsection{Data Integration System and Knowledge Graph Creation}
\noindent A data integration system can be defined as $DIS = \langle O,S,M,F \rangle$ where $O$ represents the ontology that comprises classes, properties, and relations, $S$ is a set of data sources, $M$ represents a set of mapping assertions, and finally, $F$ is a set of functional symbols representing built-in and user-defined functions. A knowledge graph is a directed graph generated from a data integration $DIS$ defined as $KG=(O,V,E)$ where $O$ represents the ontology, and $V$ is a set of nodes in the KG; nodes in V correspond to classes or instances of classes in O. $L$ is a set of directed labeled edges in the KG that relate nodes in $V$. Edges are labeled with properties and relations in $O$.\\ 

\noindent Considering the terminologies in RML and its extension covering the application of FnO, a.k.a. FNML, user-defined functions in $F$ represent the \verb|fnml:FunctionT|-\verb|ermMap|s which can express any data operation functions. However, built-in functions correspond to all predefined functions in [R2]RML such as \verb|rr:template|. The function $f$ in $F$ can be considered ``simple'' or ``composite''. A simple function is a \emph{term} $f(t_1,...,t_n)$ where $t_1,...,t_n$ are constant or variables. Contrary, function $f(t_1,...,t_n)$ is composite, if any $t_i$ is also a function. 

\subsection{Mapping Assertions}

\begin{figure*}[t!]
\centering
\includegraphics[width=1.0\linewidth]{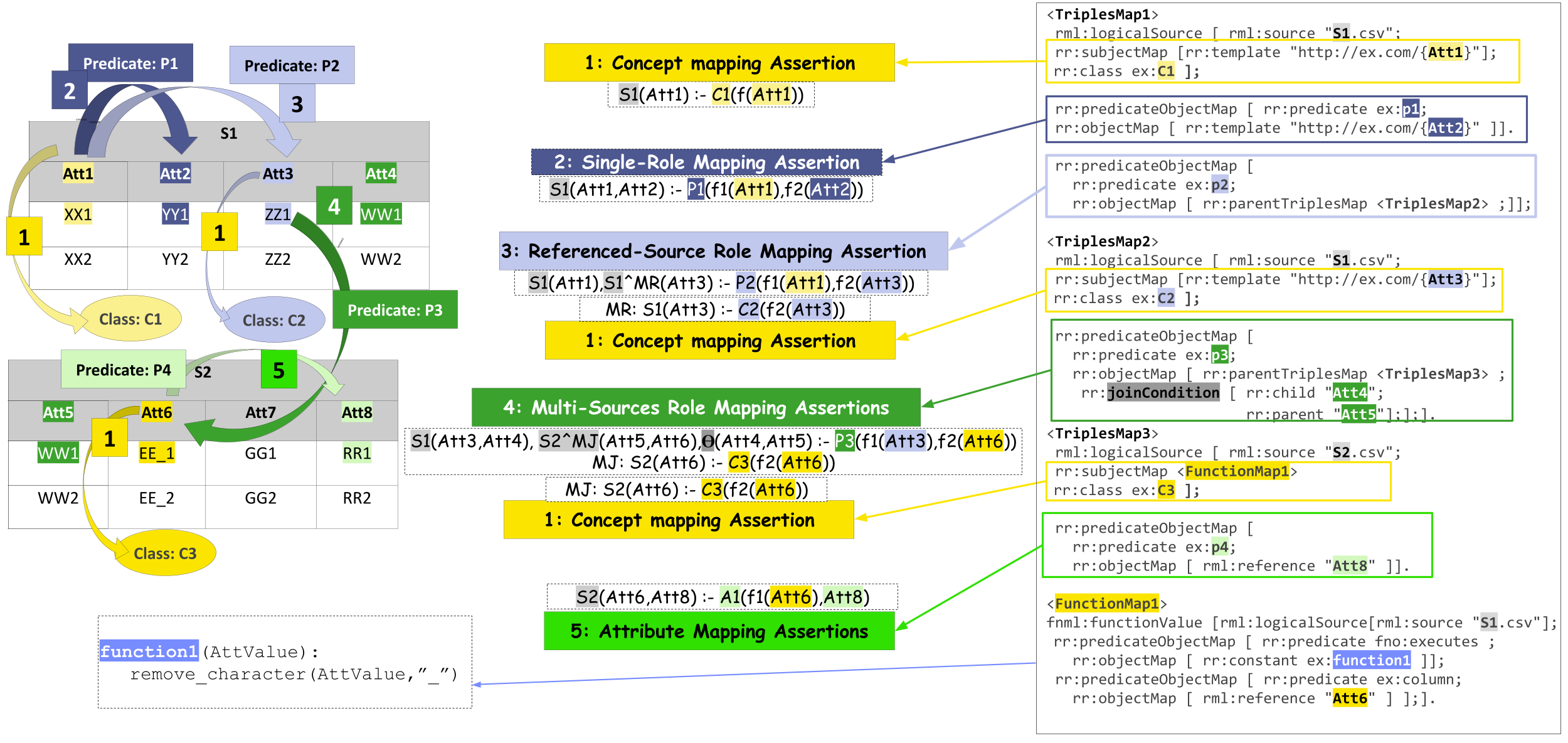}
\caption{\textbf{Mapping Assertions Illustration}.}
\label{fig:mappingExample}
\end{figure*}

\noindent Mapping rules or assertions in $M$ are formalized as definite clauses, where $body(\overline{X})$ is a conjunction of predicates over the sources in $S$. The $head(\overline{Y})$ is defined as an n-ary predicate representing classes and properties in $O$, over a set of \emph{terms}. In the following, we first provide the formal definition of different mapping assertions followed by examples illustrated in \autoref{fig:mappingExample}. Lastly, the syntax of the RML mapping language representing different mapping assertions is explained with examples shown in \autoref{fig:mappingExample}. The concept of mapping assertion extends the formalization presented in ~\cite{abs-2201-09694}.  
\\

\noindent\textbf{Concept Mapping Assertions:} define instances of classes $C$ in an ontology $O$ using the class predicate $C(.)$ over the results of the $f(.)$ receiving a \emph{term} $t$ as input arguments. The predicate $S(\overline{X})$ represents the conjunction of source signatures $S_1(\overline{X_1}),...,S_k(\overline{X_k})$, and $\overline{X}$ corresponds to the set of all the variables in the mapping assertion $m$, i.e., $\overline{X}$ is the union of $\overline{X_1},...,\overline{X_k}$. \[S(\overline{X}) :- C(f(t))\]     
\noindent In \autoref{fig:mappingExample}, two examples of concept mapping assertions are shown in \textcolor[rgb]{0.98,0.89,0.003}{\textbf{yellow}} which define the instances of the classes \verb|C1|, \verb|C2|, and \verb|C3| according to the values of the attributes in data sources; attributes \verb|Att1| and \verb|Att3| in \verb|S1|, and attribute \verb|Att6| in \verb|S2|. According to the [R2]RML terminologies, the concept mapping assertion corresponds to the \verb|rr:subjectMap| where the attributes of the \verb|rml:logicalSource|, which represents data sources, define the instances of the classes in $O$. As illustrated with the example in \autoref{fig:mappingExample}, $f(.)$ corresponds to a built-in function, represented by RDF predicate \verb|rr:template| to enable the concatenation of strings, or user-defined functions, defined by \verb|fnml:FunctionTermMap|. As shown in \autoref{fig:mappingExample}, \verb|fnml:FunctionTermMap| declares the definition of data operation functions (\verb|function1|) over attributes of the \verb|rml:logicalSource| (\verb|Att6|).\\

\noindent\textbf{Role Mapping Assertions:} defines an instance of the properties in $O$ between instances of two classes in $O$ with the role predicate $P(.)$ over the data sources attributes. This mapping assertion is expressed by \verb|predicateobjectMap| in [R2]RML. Role mapping assertion can be divided into three categories as the following. \\

\noindent\textbf{Single-Role Mapping Assertions} define $P(.,.)$ over the attributes of a single data source using $f_1(t_1),f_2(t_2)$ where $f_1$ and $f_2$ are in $F$ and $t_1$ and $t_2$ are \emph{term}s. An example of this type of role mapping assertion is shown in \autoref{fig:mappingExample} in \textcolor[rgb]{0.30,0.34,0.56}{\textbf{purple}}, where it defines the role predicate $P1(.,.)$ over the instances of the class \verb|C1| using \verb|Att1| in \verb|S1| processed by the built-in functions $f1$, and the object values based on \verb|Att2| in the same data source \verb|S1| processed by the $f2$. Using the syntax of [R2]RML, single-role mapping assertion is expressed by \verb|rr:objectMap|. \[S(\overline{X}) :- P(\underbrace{f_1(t_1)}_{\textit{Cma}},f_2(t_2))
\ \ \ \ \ \rightarrow \ \ \ \ \  Cma:\ S_1(\overline{X}_{1,1}):-C_k(f_1(t_1)))\]

\noindent\textbf{Referenced-Source Role Mapping Assertions}, similar to the previous assertion, a referenced-source role mapping assertion defines the instances of the properties in $O$ between instances of two classes in $O$ with values over the same conjunction of source signatures $S(\overline{X})$. However, this assertion allows defining the object over a term $t_2$, and a function, $f_2(.)$, which are utilized in another mapping assertion, $M$, to specify the instance of a class in $O$: 
\[S_i(\overline{X}_{i,1}),S_i^{MR}(\overline{X}_{i,2}) :- P(f_1(t_1),f_2(t_2)) \ \ ,\ \ MR: S_i(\overline{X}_{i,2}):-C_j(f_2(t_2))\] 
\noindent An example of referenced-source role mapping assertion is provided in \textcolor[rgb]{0.75,0.78,0.93}{\textbf{violet}} where the predicate $P2$ is defined over the instances of the class \verb|C1| using \verb|Att1| and the instances of the class \verb|C2| based on \verb|Att3| which are characterized as another concept mapping assertion. In [R2]RML reference-source role mapping assertions are expressed by using \verb|rr:parentTriplesMap| to reference the \verb|objectMap| of one \verb|triplesMap| (\verb|TriplesMap1| in the \autoref{fig:mappingExample}) to the \verb|subjectMap| of another \verb|triplesMap| (\verb|TriplesMap2| in the \autoref{fig:mappingExample}).\\ 

\noindent\textbf{Multi-Sources Role Mapping Assertions:} Contrary to the previous assertion, a multi-source role mapping assertion allows for expressing the instances of the properties in $O$ between instances of two classes in $O$ with values over two different sources. Since the sources $S_i$ and $S_j$ are different, a join condition. 
\[S_i(\overline{X}_{i,1}),S_j^{MJ}(\overline{X}_{i,2}),\theta (\overline{X}_{i,1},\overline{X}_{i,2}) :- P(f_1(t_1),f_2(t_2)) \ \ ,\ \ MJ: S_j(\overline{X}_{i,2}):-C_z(f_2(t_2))\] 

\noindent In \autoref{fig:mappingExample}, an example of multi-sources role mapping assertion is shown in \textcolor[rgb]{0.23,0.63,0.18}{\textbf{dark green}}. This mapping assertion defines the predicate \verb|P3| over the instances of the class \verb|C1| using \verb|Att3| in data source \verb|S1| and the instances of the class \verb|C3| based on the values in \verb|Att6| in the data source \verb|S2|. The entries of two data sources \verb|S1| and \verb|S2| are connected by the join $\theta$ between the common fields in both data sources, i.e., \verb|Att4| and \verb|Att5|. Considering the [R2]RML language, similar to the referenced-source role mapping assertion, this assertion is expressed applying \verb|rr:parentTriplesMap| connecting two \verb|triplesMap|s (\verb|TripelsMap1| and \verb|TripelsMap2|). Additionally, the join condition between two \verb|logicalSource|s is defined by \verb|rr:joinCondition| where \verb|rr:child| represents the attribute in the \verb|logicalSource| of the (\verb|triplesMap|) referencing to another (\verb|triplesMap|) as the object (\verb|TriplesMap1| in this example). Also, \verb|rr:parent| refers to the attribute in the \verb|logicalSource| of the (\verb|triplesMap|) that is referred to by \verb|rr:parentTriplesMap| (\verb|TriplesMap3| in \autoref{fig:mappingExample}).\\

\noindent\textbf{Attribute Mapping Assertions:} defines the properties of a class in $O$ using a predicate $A(.)$ over the values of attributes in terms of the conjunction of source signatures in $S(\overline{X})$. 
\[S(\overline{X}):-A(f(t_1),t_2)\]
\noindent In \autoref{fig:mappingExample} an example of attribute mapping assertion is presented (in \textcolor[rgb]{0.18,0.95,0}{\textbf{green}}) which defines the predicate $A1$ over the instances of the class \verb|C3| using \verb|Att6| in \verb|S1| and the literal data values in \verb|Att8| in the same source. R2RML provides \verb|rr:column| to represent attribute mapping assertions, while, RML introduces \verb|rml:reference|.

\begin{figure*}[!ht]
    \centering

    \subfloat[Star Join (1)]{
        \includegraphics[trim=0 0 0 21,clip,width=0.33\linewidth]{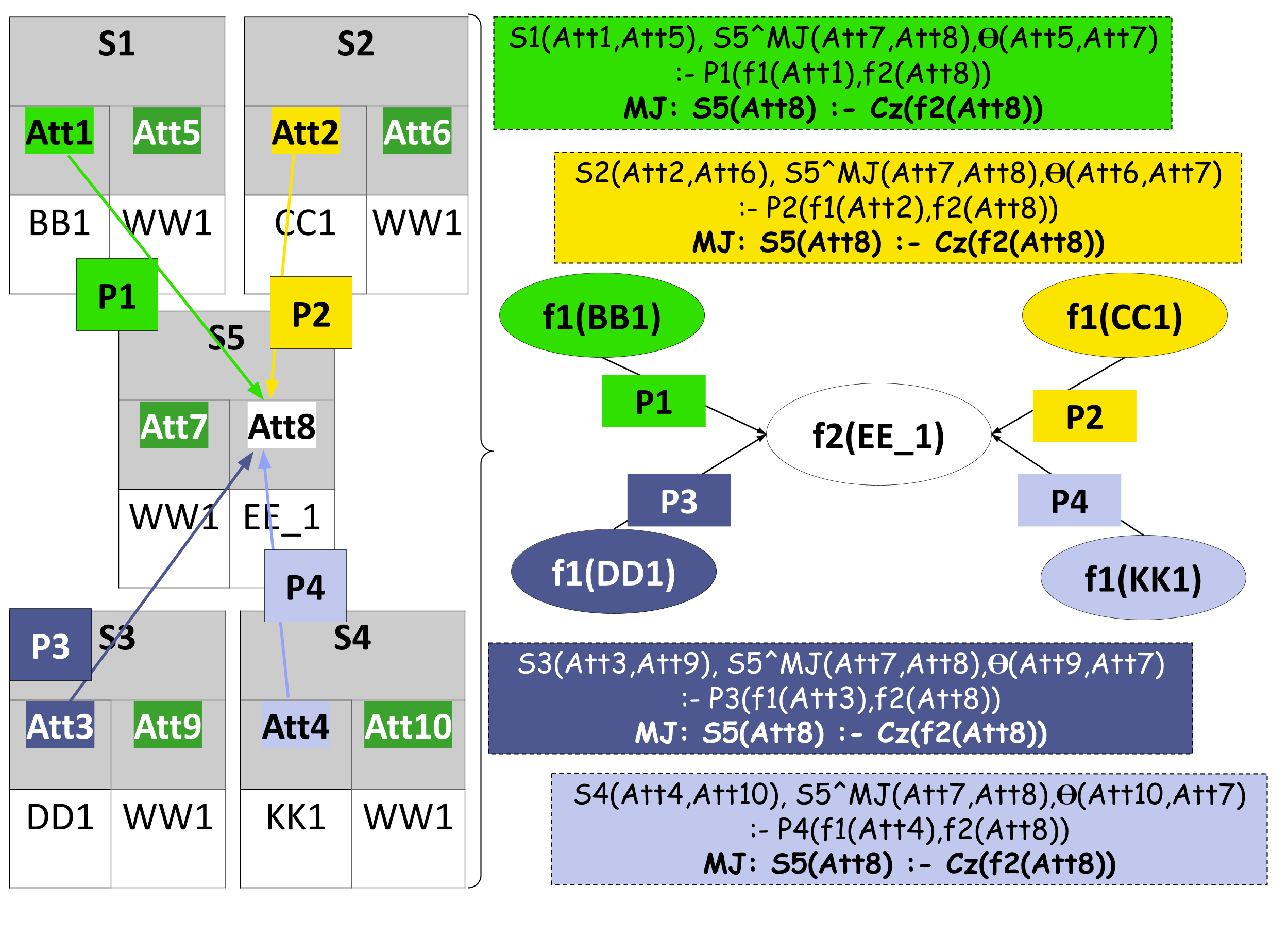}
        \label{fig:st1}
    }
    \subfloat[Star Join (2)]{
        \includegraphics[trim=0 0 0 21,clip,width=0.33\linewidth]{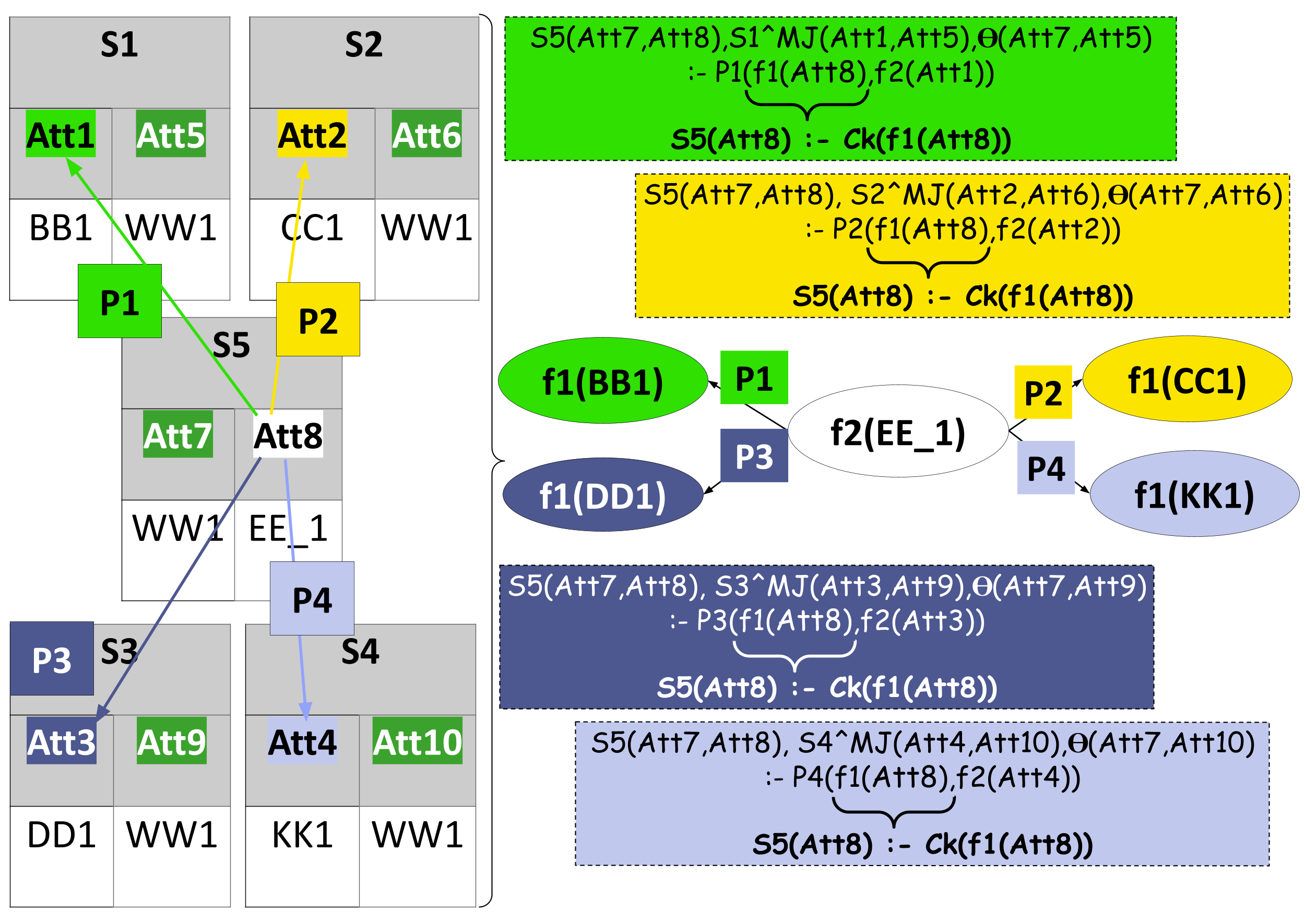}
        \label{fig:st2}
    }
    \subfloat[Chain Join]{
        \includegraphics[trim=0 0 0 21,clip,width=0.33\linewidth]{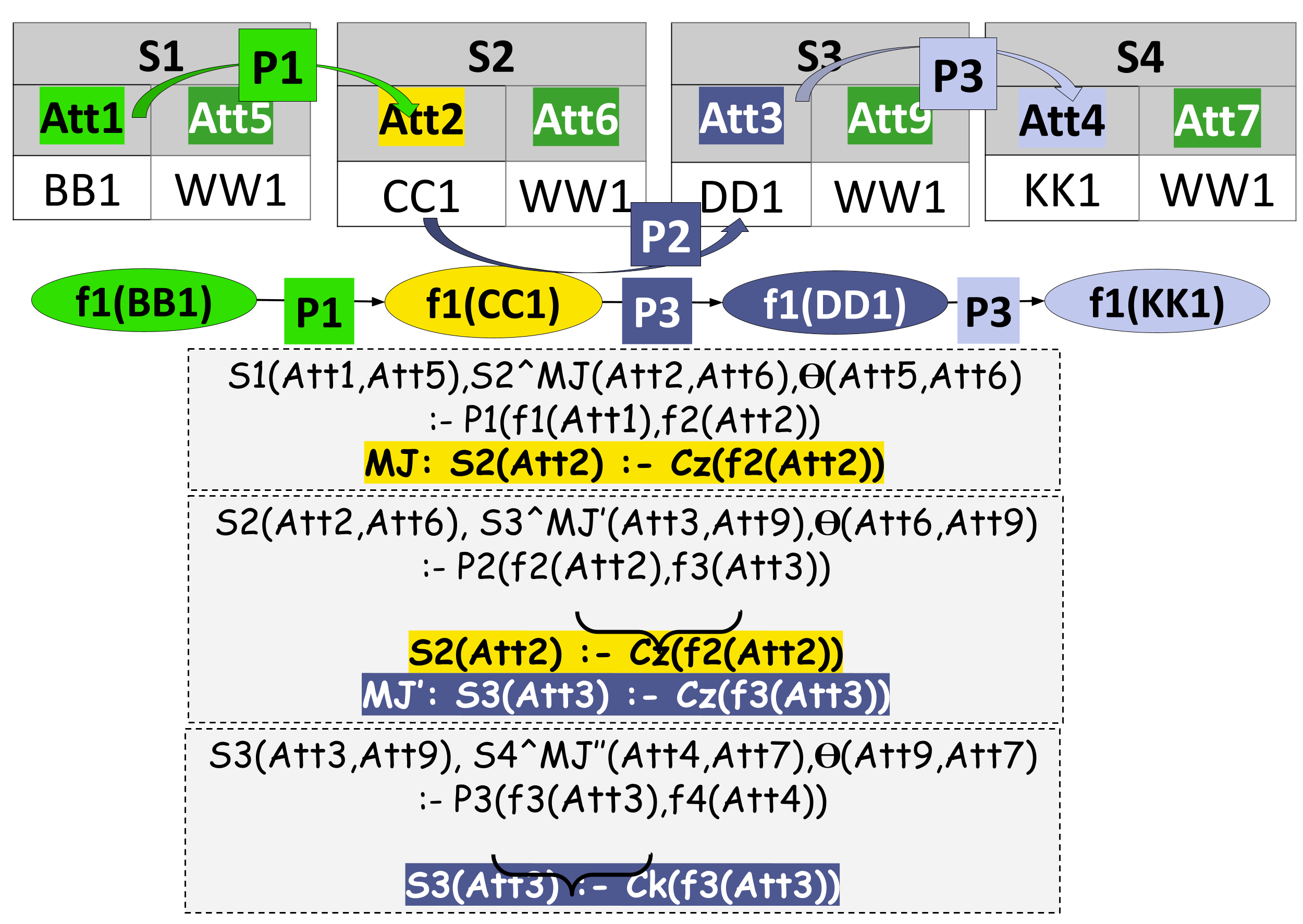}
        \label{fig:cj}
    }
    \caption{\textbf{Examples of Different Combinations of Mapping Assertions.}}
    \label{fig:star_chain_join}
\end{figure*}

\subsubsection{Combinations of Mapping Assertions}
\noindent Given a group of mapping assertions, specifically the one including more than one multi-sources role mapping assertions, they can shape expensive forms in terms of execution. Here we define two of such forms called the ``star join'' and the ``chain join''. A group of multi-sources role mapping assertions form \emph{star join}s if: \textbf{1.} more than one multi-sources mapping assertions include the same $MJ: S_j(\overline{X}_{i,2}):-C_z(f_2(t_2))$, or \textbf{2.} more than one multi-sources role mapping assertions include the same concept mapping assertion $S_i(\overline{X}_{i,1}):-C_k(f_1(t_1))$. To clarify, Figures \ref{fig:st1} and \ref{fig:st2}, each demonstrates one example of star joins corresponding to one of the mentioned conditions. As shown in \autoref{fig:st1}, in this example the star join is created due to the fact that four separated multi-sources role mapping assertions have the same definition of $MJ: S_5(Att_8):-C_z(f_2(Att_8))$, i.e., refer to the same object. However, the four different multi-sources role mapping assertions in \autoref{fig:st2} generate a start join due to the fact that they all share the same concept mapping assertion, i.e., $S_5(Att_8):-C_k(f_1(Att_8))$. Additionally, given a group of mapping assertions that is comprised of more than one multi-sources role mapping assertion, they generate a chain join if the definition of $MJ$ in at least one of the multi-sources role mapping assertions is the same as the definition of another multi-sources mapping assertions. To better understand, \autoref{fig:cj} depicts an example in which four different multi-sources role mapping assertions create a chain join. As it can be observed in \autoref{fig:cj}, the definition of $MJ$ in the first multi-sources role mapping assertion (the top one) is the same as the definition of concept mapping assertion in the second multi-sources role mapping assertion, i.e., $S_2(Att_2):-C_z(f_2(Att_2))$. Furthermore, the definition of $MJ'$ in the second multi-sources role mapping assertion is the same as the concept mapping assertion in the last multi-sources role mapping assertion, i.e., $S_3(Att_3):-C_k(F_3(Att_3))$.

\subsection{Semantics of Mapping Assertions}
\noindent This section presents the formal semantics of the mapping assertion based on model theory. First, we define an interpretation structure of a data integration system $DIS= \langle O,S,M,F \rangle$, and the interpretation of the data sources in $S$ in $I$, as well as the meaning of interpreting each of the mapping assertions of $M$ in $I$.\\

\noindent\textbf{Interpretation structure for a Data Integration System $DIS= \langle O,S,M,F \rangle$}

\noindent Let $DIS= \langle O,S,M,F \rangle$ be a data integration system, an interpretation structure $I$ for $DIS$ is defined as follows:
$I= \langle D,\sigma_c, \sigma_S,\sigma_F \rangle$, where

\begin{enumerate}
    \item $D$ is a non-empty set called domain of discourse
    \item $\sigma_c$: for each constant $c$ in $O$, $S$, $M$, and $F$, $\sigma_c(c)$ assigns a value in $D$ for $c$
    \item $\sigma_S$: for each source signature $s$ in $S$,  $\sigma_S(s)$ provides an interpretation of $s$ in $D$. If $s$ is an n-arity predicate, $\sigma_S(s)$ is a subset of $D^n$ 
    \item $\sigma_F$: for each functional symbol $f$ in $F$,  $\sigma_F(f)$ provides the interpretation of $f$ in $D$. If $f$ is an n-arity function, $\sigma_F(f): D^n->D$
\end{enumerate}

\begin{figure*}[t!]
\centering
\includegraphics[width=1.0\linewidth]{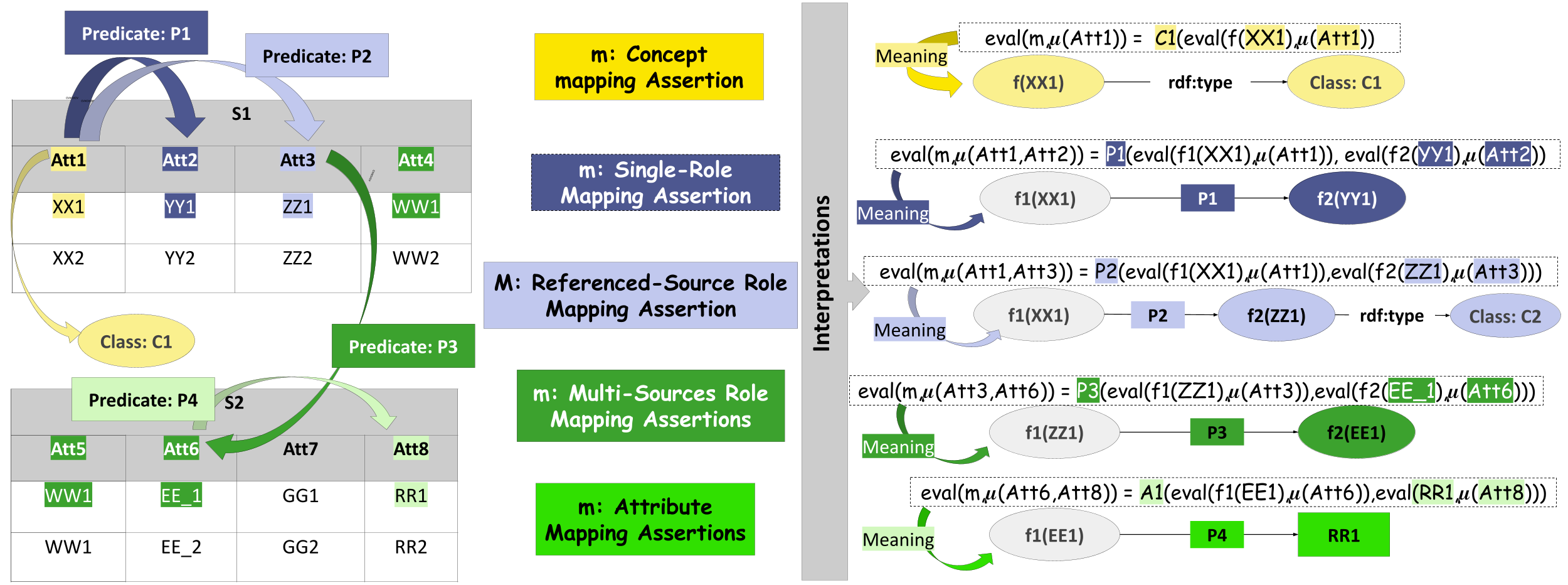}
\caption{\textbf{Mapping Assertion Interpretations}.}
\label{fig:interpretation}
\end{figure*}

\noindent\textbf{Interpretation of Sources in a Data Integration System $DIS= \langle O,S,M,F \rangle$}.
\noindent Let $S$ be a set of sources $S=\{S_1,S_2,...,S_n\}$ in a data integration system $DIS= \langle O,S,M,F \rangle$. Let $I= \langle D,\sigma_c,\sigma_S,\sigma_F \rangle$ be an interpretation structure for $DIS$. The extension of the sources in $S$ according to $I$, a.k.a. $E(S,I)$, is defined as follows:  
\begin{inparaenum}

\item $E:P^S \times \overline{I} ->  P^{\overline{S}}$, where $\overline{I}$ is a set of interpretation structures of $DIS$; and $\overline{S}$ represents an infinite set of interpretations of source signatures from $S$ according to interpretation structures in $\overline{I}$. 

\item $E(S,I)=\{\sigma_S(S_1),\sigma_S(S_2),...,\sigma_S(S_n)\}$, i.e., $E(S,I)$ returns for each of the source signature $S_i$ in $S$, a set that corresponds to the interpretation of $S_i$ in $I$.
\end{inparaenum}
\\

\noindent\textbf{Evaluation of Mapping Assertions: Interpretation of \emph{term}s in mapping assertions}. Let $I= \langle D,\sigma_c,\sigma_S,\sigma_F \rangle$ be an interpretation structure for $DIS$. Let $t$ be a term in mapping assertion $m$ in $M$. Let $\mu(\overline{X})$ be an assignment of the variables in $t$. The evaluation of the term $t$ in $\mu(\overline{X})$ according to $I$,  a.k.a. $eval(t,\mu(\overline{X}))$, is defined inductively as follows:
\textbf{Base case:} a. The term $t$ is a constant, $eval(t,\mu(\overline{X})= \sigma_c(t)$ b. The term $t$ is a variable and $\mu(\overline{X})|t$ is the value of $t$ in $\mu(\overline{X})$, $eval(t,\mu(\overline{X})=\sigma_c(\mu(\overline{X})|t)$. 
\textbf{Inductive case:}
c. If $h$ is an n-ary function and $t_1, t_2,..., t_n$ are terms, and $t=h(t_1,t_2,...,t_n)$. Then, 
$eval(h(t_1,t_2,...,t_n),\mu(\overline{X})= \sigma_F(eval(t_1,\mu(\overline{X}), eval(t_2,\mu(\overline{X}),...,eval(t_n,\mu(\overline{X}))$
\\
\\
\noindent\textbf{Interpretation of a Concept Mapping Assertion}. \\
\noindent Let $S(\overline{X}):-C(f(t))$ be a concept mapping assertion in $M$, where $S(\overline{X})$ represents the conjunction of source signatures $S_1(\overline{X_1}),...,S_k(\overline{X_k})$, and $\overline{X}$ corresponds to the set of all the variables in the mapping assertion $m$, i.e., $\overline{X}$ is the union of $\overline{X_1},...,\overline{X_k}$. Let $I= \langle D,\sigma_c,\sigma_S,\sigma_F \rangle$ be an interpretation structure for $DIS$. Let $\mu(\overline{X})$ be an assignment of the variables in $\overline{X}$ to values in $D$, such that, each $S_i(\mu(\overline{X_i}))$ belongs to $\sigma_S(S_i)$ and the interpretation of $f(t)$ in $\mu(\overline{X})$, a.k.a. $eval(f(t),\mu(\overline{X}))$ belongs to $\sigma_F(f)$. Then, $eval(m,\mu(\overline{X}))=C(eval(f(t),\mu(\overline{X})))$. To simplify the perception, we illustrate the interpretation of a concept mapping assertion with an example shown in \autoref{fig:interpretation} in \textcolor[rgb]{0.98,0.89,0.003}{\textbf{yellow}}. As it is shown in \autoref{fig:interpretation}, given a concept mapping assertion as the $m$, the interpretation of $m$ given \verb|Att1|, i.e., $eval(m,\mu (Att1))$, provides an RDF triple representing an instance of the class \verb|C1| in $O$ with the value $f(Att1)$.     
\\
\\
\noindent\textbf{Interpretation of a Single-Role Mapping Assertion}.\\
\noindent Let $S(\overline{X}):-P(f_1(t_1),f_2(t_2))$ be a single-role mapping assertion in $M$, where $S(\overline{X})$ represents the conjunction of source signatures $S_{1}(\overline{X}_{1}), …,S_{k}(\overline{X}_{k})$, and $\overline{X}$ corresponds to the set of all the variables in the mapping assertion $m$, i.e., $\overline{X}$ is the union of $\overline{X}_{1},...,\overline{X}_{k}$. Let $I=\langle D,\sigma_c,\sigma_S,\sigma_F \rangle$ be an interpretation structure for $DIS$. Let $\mu(\overline{X})$ be an assignment of the variables in $\overline{X}$ the values in $D$, such that, each $S_{i}(\mu(\overline{X}_{i}))$ belongs to $\sigma_S(S_{})$ (for all $i$ in \{1,...,k\}) and the interpretation of $f_1(t_1)$ and $f_2(t_2)$ in $\mu(\overline{X})$, a.k.a. $eval(f_1(t_1),\mu(\overline{X}))$ and $eval(f_2(t_2),\mu(\overline{X}))$ belong to $\sigma_F(f)$. Then, $eval(m,\mu(\overline{X}))=
P(eval(f_1(t_1),\mu(\overline{X})),eval(f_2(t2),\mu(\overline{X})))$. In \autoref{fig:interpretation}, we demonstrate an example of the interpretation of a single-role mapping assertion given as $m$ in \textcolor[rgb]{0.30,0.34,0.56}{\textbf{purple}}. As it can be observed, the interpretation of $m$ given \verb|Att1| and \verb|Att2| of the same data source \verb|S1|, i.e., $eval(m,\mu(Att1,Att2))$, generates RDF triples representing the relation between the values of $f1(Att1)$ and $f2(Att2)$ defined by the predicate \verb|P1| of $O$.\\
\\
\noindent\textbf{Interpretation of a Referenced-Source Mapping Assertion}. \\
\noindent Let $S(\overline{X}_{1}), S^{MR}(\overline{X}_{2}):-P(f_1(t_1),f_2(t_2))$ be a reference-role mapping assertion in $M$ and
$MR$ is the referenced concept mapping assertion $S(\overline{X}_{2}):- C(f(t))$. The predicates $S(\overline{X}_{1})$ and $S^{MR}(\overline{X}_{2})$ represent the conjunction of source signatures $S_1(\overline{X}_{1,1}),...,S_k(\overline{X}_{1,k})$ and $S^{MR}_{1}(\overline{X}_{2,1}),...,S_k(\overline{X}_{k,2})$, respectively. Let $\overline{X}$ be the union of the variables in $\overline{X}_{1}$ and $\overline{X}_{2}$. Let $I= \langle D,\sigma_c,\sigma_S,\sigma_F \rangle$ be an interpretation structure for $DIS$. Let $\mu(\overline{X})$ be an assignment of the variables in $\overline{X}$ in $D$, such that, each $S_i(\mu(\overline{X}))$ belongs to $\sigma_S(S_i)$, and the interpretation of $f_1(t_1)$ and $f_2(t_2)$ in $\mu(\overline{X})$, a.k.a. $eval(f_j(t_j),\mu(\overline{X})$ belongs to $\sigma_F(f)$. Then, $eval(m,\mu(\overline{X}))=
P(eval(f_1(t_1),\mu(\overline{X})),eval(f_2(t2),\mu(\overline{X})))$. To clarify, an example of the interpretation of $m$ as a referenced-source mapping assertion is provided in \autoref{fig:interpretation} in \textcolor[rgb]{0.75,0.78,0.93}{\textbf{violet}}. The interpretation of $m$ given \verb|Attr1| and \verb|Attr3| of the same source \verb|S1|, generates RDF triples representing the relation between instances of two classes in $O$, i.e., \verb|C1| and \verb|C2| with the values in \verb|Attr1| and \verb|Attr3| respectively, considering the predicate \verb|P2| in $O$.\\
\\
\noindent\textbf{Interpretation of a Multi-Sources Role Mapping Assertion}.\\
\noindent Let $S(\overline{X}_{1}), S^{MJ}(\overline{X}_{2}):-P(f_1(t_1),f_2(t_2))$ be a reference-role mapping assertion in $M$ and
$MJ$ is the referenced concept mapping assertion $S^{MJ}(\overline{X}_{2}):- C_z(f(t))$. The predicates $S(\overline{X}_{1})$ and $S^{MJ}(\overline{X}_{2})$ represent the conjunction of source signatures $S_1(\overline{X}_{1,1}),...,S_k(\overline{X}_{1,k})$ and $S^{MJ}_{1}(\overline{X}_{2,1}),...,S_t(\overline{X}_{t,2})$, respectively. Let $\overline{X}$ be the union of the variables in $\overline{X}_{1}$ and $\overline{X}_{2}$. Let $I= \langle D,\sigma_c,\sigma_S,\sigma_F \rangle$ be an interpretation structure for $DIS$. Let $\mu(\overline{X})$ be an assignment of the variables in $\overline{X}$ in $D$, such that, each $S_i(\mu(\overline{X}))$ belongs to $\sigma_S(S_i)$, each a $S^{MJ}_{j}(\mu(\overline{X}))$ belongs to $\sigma_S(S^{MJ}_{j})$, and the interpretation of $f_1(t_1)$ and $f_2(t_2)$ in $\mu(\overline{X})$, a.k.a. $eval(f_j(t_j),\mu(\overline{X})$ belongs to $\sigma_F(f)$. Then, $eval(m,\mu(\overline{X}))=
P(eval(f_1(t_1),\mu(\overline{X}),$ $eval(f_2(t2),\mu(\overline{X})))$. An example of the interpretation of $m$ as a multi-sources role mapping assertion can be observed in \autoref{fig:interpretation} in \textcolor[rgb]{0.23,0.63,0.18}{\textbf{dark green}}. Similar to the interpretation of referenced-role mapping assertion, the interpretation of multi-sources role mapping assertion produce RDF triples representing the relation between instances of two classes in the $O$, i.e., \verb|C1|and \verb|C3|, using a predicate in the $O$, i.e., \verb|P3| in this example. What differentiate the two interpretations are the data sources which contribute in the creation of the instances of \verb|C1| and \verb|C3|. As it can be seen in \autoref{fig:interpretation}, \verb|Attr1| in the data source \verb|S1| provides the instantiation of the class \verb|C1|. However, the instances of the class \verb|C3| are provided by the values of \verb|Attr6| in the other data source, i.e., \verb|S2|. \\
\\
\noindent\textbf{Interpretation of an Attribute Mapping Assertion}.\\
\noindent Let $S(\overline{X}):-A(f_1(t_1),f_2(t_2))$ be an attribute mapping assertion in $M$, where $S(\overline{X})$ represents the conjunction of source signatures $S_{1}(\overline{X}_{1}), …,S_{k}(\overline{X}_{k})$, and $\overline{X}$ corresponds to the set of all the variables in the mapping assertion $m$, i.e., $\overline{X}$ is the union of $\overline{X}_{1},...,\overline{X}_{k}$. Let $I=\langle D,\sigma_c,\sigma_S,\sigma_F \rangle$ be an interpretation structure for $DIS$. Let $\mu(\overline{X})$ be an assignment of the variables in $\overline{X}$ the values in $D$, such that, each $S_{i}(\mu(\overline{X}_{i}))$ belongs to $\sigma_S(S_{})$ (for all $i$ in \{1,...,k \}) and the interpretation of $f_(t_1)$, a.k.a. $eval(f_1(t_1),\mu(\overline{X}))$, belongs to $\sigma_F(f)$ and the interpretation of $t_2$ in $\mu(\overline{X})$, a.k.a. $eval(t_2,\mu(\overline{X}))$, is true in $I$. Then, $eval(m,\mu(\overline{X}))=
P(eval(f_1(t_1),\mu(\overline{X})),eval(t2,\mu(\overline{X})))$. The RDF triples generated by the interpretation of attribute mapping assertions represent the relation between instances of a class in $O$ and a literal value extracted from an attribute in the same data source using a predicate in the $O$. In \autoref{fig:interpretation} an example of the interpretation of an attribute mapping assertion is provided in \textcolor[rgb]{0.18,0.95,0}{\textbf{green}}. It shows the relation between the instances of the class \verb|C3| and literal values extracted from the \verb|Att8| in \verb|S2| using the predicate \verb|P4|.\\
\\
\noindent 
\noindent\textbf{Evaluation of a Data Integration System in an Interpretation Structure.}
Let $DIS= \langle O,S,M,F \rangle$ be a data integration system. Let $I= \langle D,\sigma_c,\sigma_S,\sigma_F \rangle$ be an interpretation structure for $DIS$. The evaluation of $DIS$ in $I$, a.k.a. $evaluation(DIS,I)$, corresponds to the union of $eval(m,\mu(\overline{X}))$ for all the mapping assertions in $M$ and all the assignments $\mu(\overline{X})$ in $D$ of the variables $\overline{X}$ in $m$. The evaluation of $DIS$ in $I$, $evaluation(DIS,I)$ denotes the instances of knowledge graph $KG=(O,V,E)$. 

\begin{figure}[t!]
\centering
\includegraphics[trim=0 0 0 21,clip,width=0.8\linewidth]{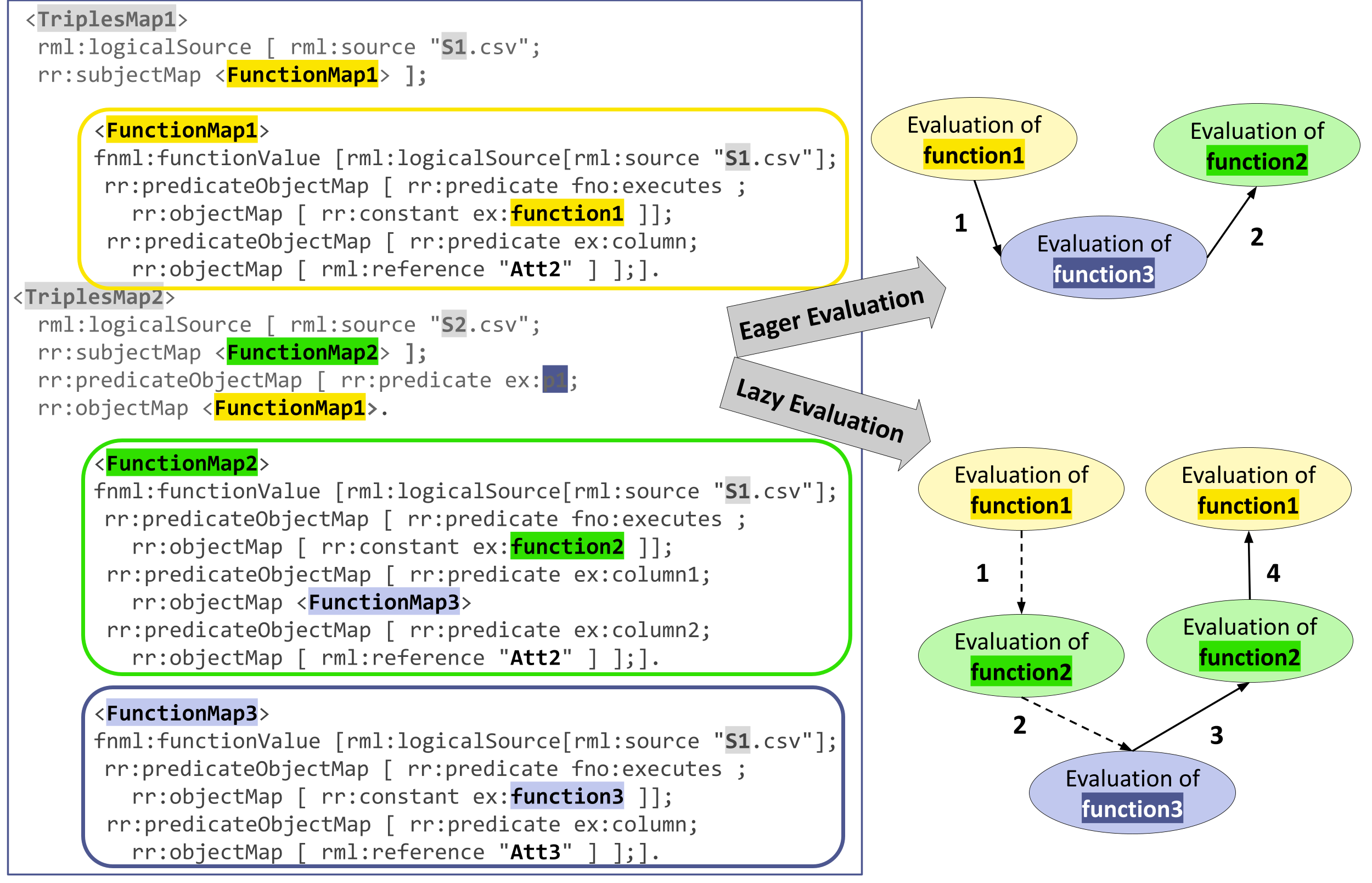}
\caption{\textbf{Evaluation Strategies Example}. The two evaluation approaches including eager and lazy evaluations are shown for the same set of functions. Following an eager evaluation strategy, the function that is repeated in two mapping assertions, i.e., \textit{function1} is evaluated only once. Also, \textit{function3} which is an argument to \textit{function2} is evaluated prior to \textit{function2}. In contrast, following a lazy evaluation strategy, the evaluation of \textit{function1} is repeated as many times as it is referred to in mapping assertions. As \textit{function2} appears in mapping assertions prior to \textit{function3}, a lazy evaluation strategy tries to evaluate \textit{function2} prior to \textit{function3} which cannot be completed and is revisited after the evaluation of \textit{function3}.}
\label{fig:ev_example}
\end{figure}

\subsection{Evaluation of Functions}
\noindent The interpretation of $f(t)$ in $\mu(\overline{X})$, i.e., $eval(f(t),\mu(\overline{X}))$ can be derived using a lazy or an eager evaluation approach. The evaluation of a function is called eager when the parameters of the function are evaluated before the function is executed~\cite{torra2016lazy}. Moreover, employing an eager evaluation strategy means that functions, $f(t) \in F$, are executed ``as soon as possible''. In contrast, a lazy evaluation approach leads to evaluating functions ``as required''. For the sake of understandability, we explain the two strategies of the eager and lazy evaluation with the exemplar mapping assertions shown in \autoref{fig:ev_example}. The intuitive first step of an eager-evaluation-based approach is to traverse all the mapping assertions, find the ones including user-defined functions, and evaluate the functions. Accordingly, having the mapping assertions shown in \autoref{fig:ev_example}, such an eager-evaluation-based approach, first, detects four mapping assertions including user-defined functions. Then it starts evaluating them. However, since two of the functions are the same, the eager evaluation enables the approach to evaluate the duplicated function exactly once. Moreover, considering the fact that one of the arguments of the \verb|function2| is the output of the \verb|function3|, the eager evaluation forces the evaluation of the \verb|function3| prior to the evaluation of the \verb|function2|. In contrast, a lazy-evaluation-based approach starts executing the functions as soon as it reads the corresponding mapping assertions. One of the drawbacks of such an approach is that the duplicated functions are evaluated multiple times.

\section{Our Proposed Approach: Dragoman}
\label{sec:approach}
\noindent 
This section defines the problem tackled in this paper, and the solution proposed to efficiently create knowledge graphs. 
As formally shown in the previous section, a knowledge graph
corresponds to the evaluation of a data integration system, $DIS_G = \langle O,S,M,F \rangle$ in an interpretation structure $I$. The mapping assertions in $M$ state the definition of the concepts in the ontology $O$ in terms of the data source signatures in $S$. Following existing W3C standards, these mapping assertions can be specified in R2RML and RML. $F$ is a set of built-in and user-defined functions for data operations, presented declaratively applying languages such as FnO+RML.
This paper addresses the problem of evaluating the user-defined functions efficiently and optimizing the evaluation of the given DIS. Dragoman, the proposed solution in this paper, interprets and transforms the mapping assertions in $M$ and evaluates the functions in $F$ based on the data in $S$ efficiently. Dragoman focuses on scaling up the function evaluation process and transformation of a provided $DIS$ into an optimized, function-free one, as part of the knowledge graph creation process. The outcome of Dragoman is a transformed data integration system as $DIS_G = \langle O,S',M' \rangle$.

\begin{figure}[!ht]
\centering
\includegraphics[width=0.8\linewidth]{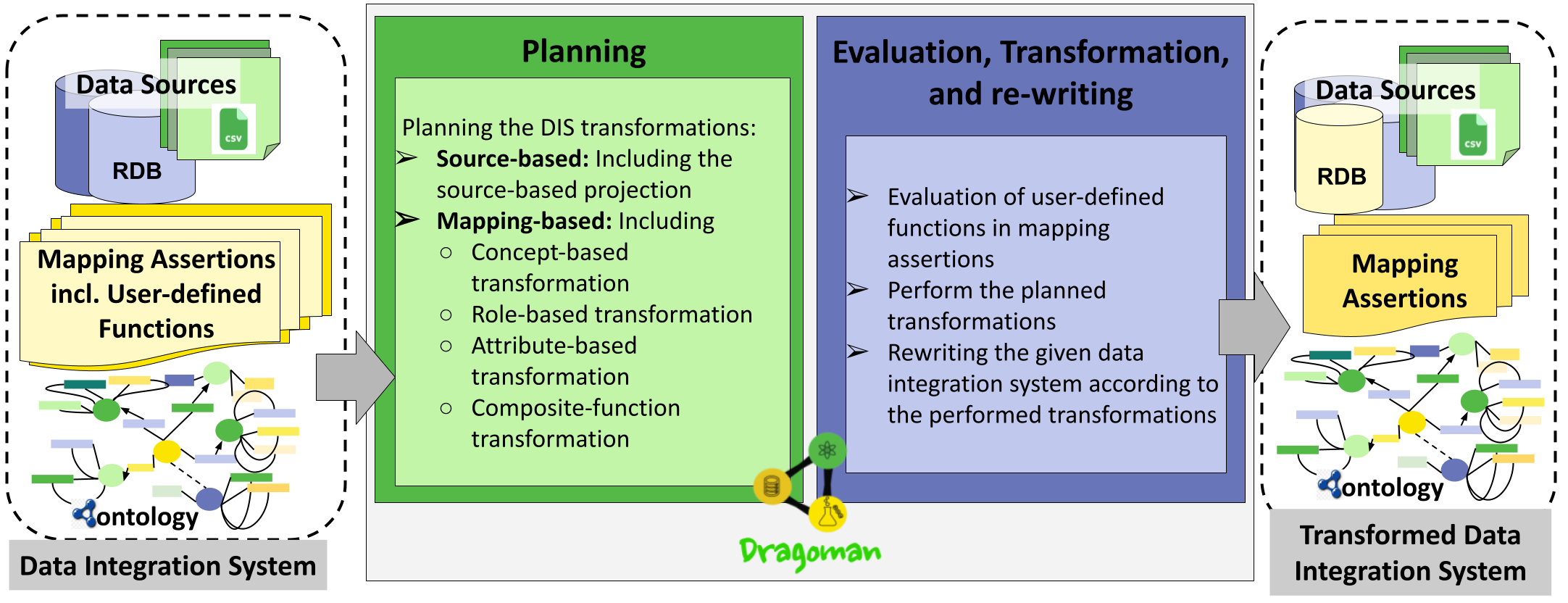}
\caption{\textbf{Approach}. Dragoman framework provides a set of transformation rules divided into source-based and mapping-based rules. First, considering the given DIS, Dragoman plans the transformations that are required to be performed. Then, Dragoman evaluates all the user-defined functions, performs the transformations, and finally, re-writes the DIS.}
\label{fig:approach}
\end{figure}

\subsection{Problem Statement}
\noindent Given a data integration system $DIS_G = \langle O,S,M,F \rangle$ which generates the knowledge graph $G$, the problem of scaling up the process of knowledge graph creation is defined as the problem of identifying a data integration system $DIS'_G = \langle O,S',M',F \rangle$ such that: \textbf{i)} The execution time of $eval(m',\mu(\overline{X'}))$ is \emph{minimized} for all $m' \in M'$ and $\overline{X'} \in S'$.
\textbf{b)} The RDF knowledge graphs resulted from evaluating the two data integration systems are \emph{equivalent}, i.e., $\Sigma^n_{i=1} eval(m_i,\mu(\overline{X_i}))=\Sigma^{n'}_{j=1}eval({m'}_j,\mu(\overline{X'_j}))$.

\subsection{Proposed Solution}
\noindent We propose Dragoman, to efficiently evaluating \underline{d}ecla\underline{ra}tive mappin\underline{g} languages \underline{o}ver fra\underline{m}eworks for knowledge gr\underline{a}ph creatio\underline{n}. Dragoman framework introduces a set of transformation rules to transform a given $DIS_G$ to the $DIS'_G$ such that the execution time required to create the same knowledge graph $G$ from $DIS'_G$ is less than the required time by $DIS$ to produce the same knowledge graph $G$. As shown in the \autoref{fig:approach}, the transformation rules proposed by Dragoman are grouped into source-based and mapping-based categories. Dragoman plans the required transformations based on the mapping assertions in $M$ given by $DIS_G$. We explain the transformation rules in details in section \ref{sec:transformations}.

\begin{algorithm}[H]
\centering
\caption{Evaluation, Transformation, and re-writing Algorithm}
\label{alg:algo1}
\begin{multicols}{2}
\begin{algorithmic}[1]
\Require $ OLD\_DIS = \langle O,S,M,F \rangle$ and
$TRs$
\Ensure $NEW\_DIS = \langle O,S',M',F \rangle$
\State Suppose $OLD\_DIS.\overline{MF}$ is a set of mapping assertions including \emph{term}s $g_i(.)$ that are user-defined functions: $OLD\_DIS.F:\{g_1(.),...,g_n(.)\}$ and receive a set of \emph{term}s $\overline{T}_i$ as the arguments.
\Repeat 
\State Select from $OLD\_DIS.M$ a mapping 
\State assertion $ma$
\If{ $ma$ in $OLD\_DIS.\overline{MF}$}
\State $S_g(\overline{X}_g,\overline{T}) =$ \State $Execute(S_{ma}(\overline{X}_{ma}), g_{ma}(\overline{T}_{ma}),$ $OLD\_DIS.F)$
\EndIf
\State $(\overline{Tma},\overline{TS}) \gets Transformation($ $ma,TRs)$
\State $M' = (OLD\_DIS.M - \{ma\}) \cup \overline{Tma}$
\State $S' = (OLD\_DIS.S \cup \overline{TS})$
\State $NEW\_DIS.M = M'$
\State $NEW_DIS.S = S'$
\Until{$OLD\_DIS==NEW\_DIS$} \\
\Return $NEW\_DIS$
\end{algorithmic}
\end{multicols}
\end{algorithm}


\noindent Algorithm\ref{alg:algo1} represents the evaluation, transformation, and re-writing component in Dragoman, also shown in \autoref{fig:approach}. As it can be perceived from the Algorithm\ref{alg:algo1}, Dragoman relies on an eager evaluation strategy in evaluating user-defined functions in $F$. Intuitively, the eager evaluation of user-defined functions is the first step in the execution algorithm of Dragoman (lines 1-8). Algorithm\ref{alg:algo_exe} expresses a sketch of the steps that Dragoman follows in evaluating user-defined functions. To meet the requirements of an eager evaluation, in the case of having composite functions, the priority of the execution is with the evaluation of functions which are the arguments to the other functions; Algorithm\ref{alg:algo_exe} lines 2 and 3. The Eager evaluation strategy enables Dragoman to avoid evaluating the same functions in $F$ with the same arguments more than once, as illustrated in the example in \autoref{fig:ev_example}.   

\begin{algorithm}[!t]
\centering
\caption{Function Evaluation Algorithm}
\label{alg:algo_exe}
\begin{multicols}{2}
\begin{algorithmic}[1]
\Require $S_{ma}(\overline{X}_{ma}),g_{ma}(\overline{T}_{ma}),OLD\_DIS.F$
\Ensure $Execute(S_{ma}(\overline{X}_{ma}),g_{ma}(\overline{T}_{ma}))$
\State $S_g(\overline{X}_g,\overline{T}) \gets \{\}$
\If{$\overline{T}_{ma}\in OLD\_DIS.F$ :  $g_{ma}(\overline{T}_{ma}) = a_{ma}(b_{ma}(\overline{T}'_{ma}))$}
\State $Execute(S_{ma}(\overline{X}_{ma}),g_{ma}(\overline{T}_{ma}),OLD\_DIS.F)$
\Else
\State $S_g(\overline{X}_g,\overline{T})$ is generated as a new data source
\State such that $\mu(t,\overline{X}_{ma})\in \overline{X}_g$ holds for each $t$  
\State in $\overline{T}_{ma}$ and $S_g(\overline{X}_g,\overline{T})$ belongs to $\sigma_S(S_g)$ 

\EndIf
\\
\Return $S_g(\overline{X}_g,\overline{T})$
\end{algorithmic}
\end{multicols}
\end{algorithm}

\subsubsection{Transformation Rules}
\label{sec:transformations}
\noindent As explained in Algorithm\ref{alg:algo1}, lines 9-14, and Algorithm\ref{alg:algo_tr}, after evaluating all the functions in $F$, Dragoman transforms the mapping assertions in $M$ and data sources in $S$ to the function-free mapping assertions in $M'$ and sources in $S'$. The key point in optimizing the process of knowledge graph creation is to consider the data sources and mapping assertions coherently~\cite{jozashoori2019mapsdi,jozashoori2020funmap}. The semantics encoded in the mapping assertions provide insight into the portions of each data source that contribute to the creation of the knowledge graph and the intersections between the data sources. Considering the different types of mapping assertions and user-defined functions, Dragoman introduces five transformation rules as described in the following.

\begin{algorithm}[H]
\centering
\caption{Transformation Algorithm}
\label{alg:algo_tr}
\begin{algorithmic}[1]
\Require $ma,\ TRs$
\Ensure $Transformation(ma,TRs)$
\State $OLD\_\overline{Tma}\gets\{\}$ 
\State $NEW\_\overline{Tma}\gets\{\}$
\State $TS\gets\{\}$
\For{each $S_i$ in $Sources(ma)$}
\State $S'_i \gets \Pi_{Att_{(ma)}}S_i$
\State $ma \gets relpace(S_i,S'_i,ma)$
\State $\overline{TS}\gets{TS}\cup\{S'_i\}$
\EndFor
\State $NEW\_\overline{Tma}\gets\{ma\}$
\Repeat
\State $OLD\_DIS \gets NEW\_DIS$
\Switch{$TR$}
\Case{$Concept-based\ Transformation$}
\State $(NEW\_\overline{Tma},\overline{TS})$ 
\State $\gets Concept\_Transformation(ma,\overline{TS})$
\EndCase
\Case{$Role-based\ Transformation$}
\State $(NEW\_\overline{Tma},\overline{TS})$ 
\State $\gets Role\_Transformation(ma,\overline{TS})$
\EndCase
\Case{$Attribute-based\ Transformation$}
\State $(NEW\_\overline{Tma},\overline{TS})$ 
\State $\gets Attribute\_Transformation(ma,\overline{TS})$
\EndCase
\EndSwitch
\Until{$OLD\_\overline{Tma}=NEW\_\overline{Tma}$}\\
\Return $OLD\_\overline{Tma},\overline{TS}$
\end{algorithmic}
\end{algorithm}

\begin{figure*}[!ht]
    \centering
    \subfloat[Source-based Projection]{
        \includegraphics[trim=0 0 0 21,clip,width=0.5\linewidth]{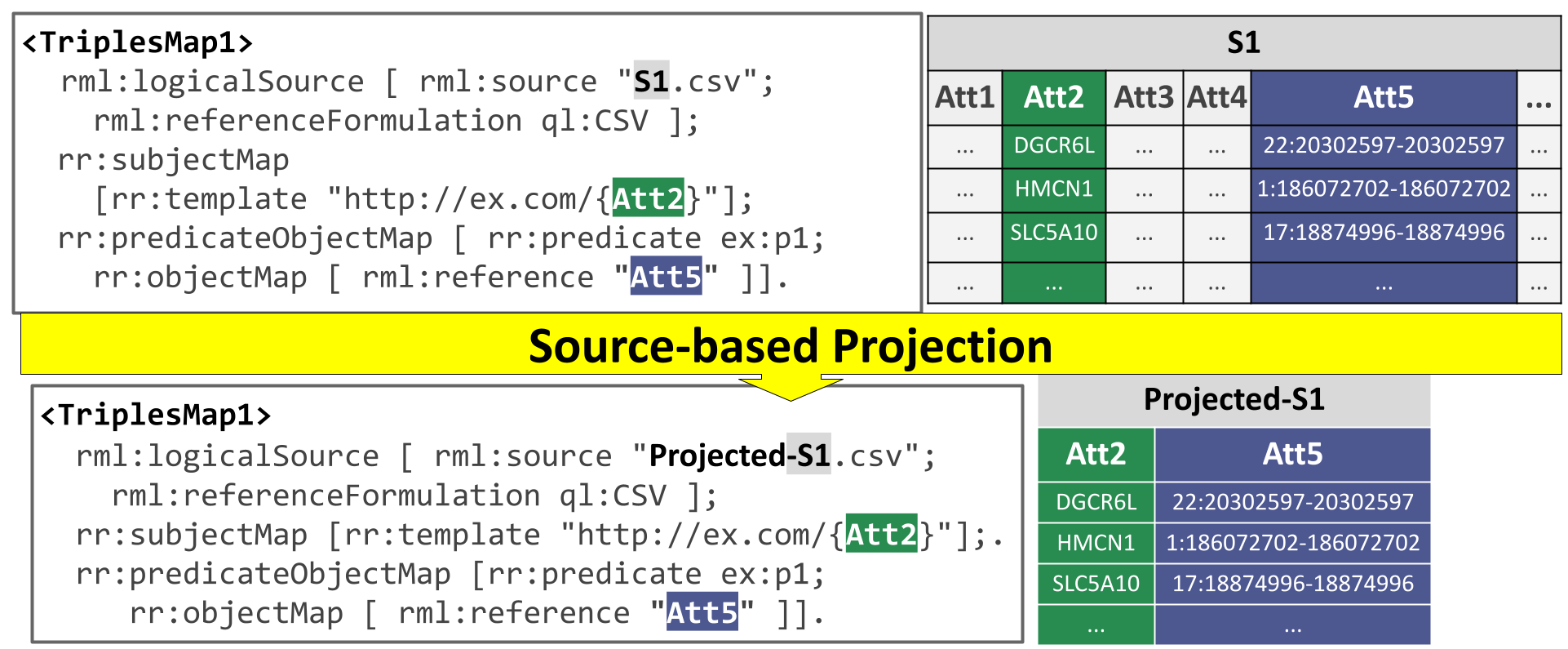}
        \label{fig:t1}
    }
    \subfloat[Concept-based Transformation]{
        \includegraphics[trim=0 0 0 21,clip,width=0.5\linewidth]{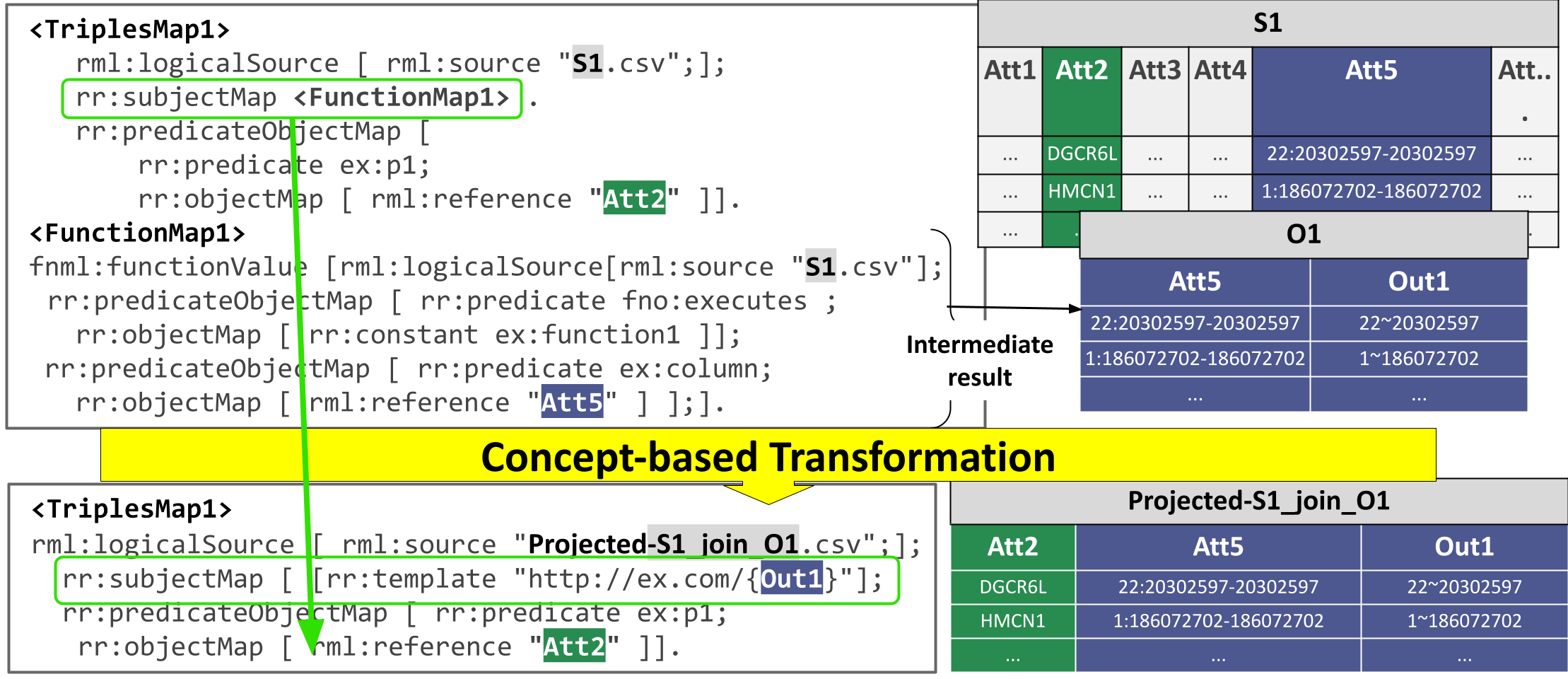}
        \label{fig:t2}
    }
    \caption{\textbf{Dragoman Source-based Projection and Concept-based Transformation}}
    \label{fig:transformations1}
\end{figure*}

\noindent\textbf{Source-based Projection.} The most generic transformation that Dragoman performs before executing the functions on any mapping assertion provided in $M$ is the projection of the attributes in sources of $S$ referred to by the mapping assertion in $M$. For each mapping assertion in $M$, independent of the type, Dragoman projects all the attributes needed by that mapping assertion into a new data source and removes the duplicated values. It should be noted that if a function in $F$ is referred to by a mapping assertion, then the attributes required are also projected into the new data source. In other words, this transformation pushes down the projection of required attributes and the duplicated values removal~\cite{jozashoori2019mapsdi}. Accordingly, the mapping assertion is transformed to use the new projected data source instead of the original one. To better understand, Figure \ref{fig:t1} shows an example of this transformation. As it can be seen in Figure \ref{fig:t1}, from the source $S1$ only two attributes $Att2$ and $Att5$ are utilized in the mapping assertions; $Att2$ as the value for \verb|rr:subjectMap| or the concept mapping assertion, and $Att5$ as the value of the \verb|rr:objectMap| or the single-role mapping assertion. Therefore, the transformed DIS (shown at the bottom of Figure \ref{fig:t1}) is comprised of the newly generated data source, including only $Att2$ and $Att5$, and the transformed mapping assertion which applies the later data source as the \verb|rr:logicalSource|. 

\noindent\textbf{Concept-based Transformation.} When the output of user-defined functions (\verb|fnml:FunctionTermMap|) is applied as the value of a concept mapping assertion or \verb|rr:subjectMap|, the concept-based transformation is performed on the given mapping assertions in $M$ and sources in $S$. Figure \ref{fig:t2} shows an example where the transformation of sources in $S$ is performed by joining the data sources consisting of the function's output, i.e., ``intermediate result'', and the outcome of the source-based projection on $S$. After the generation of the new data source, Dragoman, transforms the mapping assertion so that the old data source is replaced with the newly generated one, as shown at the bottom of Figure \ref{fig:t2}. Algorithm\ref{alg:algo_concept} provides a sketch of the concept-based Transformation rule and how to implement it. 

\begin{algorithm}[H]
\centering
\caption{Concept-based Transformation Algorithm}
\label{alg:algo_concept}
\begin{multicols}{2}
\begin{algorithmic}[1]
\Require $ma,\overline{TS}$
\Ensure $Concept\_Transformation(ma,\overline{TS})$
\State $NEW\_\overline{Tma}\gets\{ma\}$
\State $OLD\_\overline{Tma}\gets NEW\_\overline{Tma}$
\State suppose $ma: S(\overline{X}):-C(f(g(\overline{T})))$ 
\State where $g$ is a \emph{term} that is a user-defined 
\State function and receives $\overline{T}$, a set of \emph{term}s,
\State as arguments: $\overline{T}:\{t_1,...,t_m\}$, $\overline{Rma}$
\State is a set of $Rma_i$ defined as: 
\State $S_i(\overline{X}_{i,1}):-P(ma,f_2(t_2))$, $\overline{Ama}$
\State is a set of $Ama_i$ defined as:
\State $S_i(\overline{X}_{i,1}):-P(ma,t_2)$
\If {$g(.)=a(b(.))$}
\State $S_g(\overline{X}_g,b(\overline{T}))$ is defined for all \State the values of $b(\overline{T})$ such that
\State $eval(S_g(\overline{X}_g,b(\overline{T})))=True$
\State $ma':$
\State $S(\overline{X}),S_g(\overline{X}_g,b(\overline{T})):-C(f(t'))$
\State where $t'$ is a \emph{term}: $t' \in S_g$
\For{each $Rma:$ in $\overline{Rma}$}
\State $Rma':S_i(\overline{X}_{i,1}),$
\State $S_g(\overline{X}_g,b(\overline{T}):-P(ma',f_2(t_2))$
\State $\overline{Rma'}\gets \overline{Rma'}\cup Rma'$
\EndFor
\For{each $Ama:$ in $\overline{Ama}$}
\State $Ama':S_i(\overline{X}_{i,1}),$
\State $S_g(\overline{X}_g,b(\overline{T}):-P(ma',t_2)$
\State $\overline{Ama'}\gets \overline{Ama'}\cup Ama'$
\EndFor
\Else
\State $S_g(\overline{X}_g,\overline{T})$ is defined for all the
\State values of $t_i \in \overline{T}, 0<t_i<m+1$ 
\State such that $eval(S_g(\overline{X}_g,\overline{T}))=True$
\State $ma':$
\State $S(\overline{X}),S_g(\overline{X}_g,\overline{T}):-C(f(t'))$
\State where $t'$ is a \emph{term}: $t' \in S_g$
\For{each $Rma:$ in $\overline{Rma}$}
\State $Rma':S_i(\overline{X}_{i,1}),$
\State $S_g(\overline{X}_g,\overline{T}):-P(ma',f_2(t_2))$
\State $\overline{Rma'}\gets \overline{Rma'}\cup Rma'$
\EndFor
\For{each $Ama:$ in $\overline{Ama}$}
\State $Ama':\ S_i(\overline{X}_{i,1}),$
\State $S_g(\overline{X}_g,\overline{T}):-P(ma',t_2)$
\State $\overline{Ama'}\gets \overline{Ama'}\cup Ama'$
\EndFor
\EndIf
\State $NEW\_\overline{Tma}\gets \{ma'\}\cup{\overline{Rma}}\cup{\overline{Ama}}$
\State $\overline{TS}\gets\overline{TS}\cup{S_g}$\\
\Return $NEW\_\overline{Tma},\overline{TS}$
\end{algorithmic}
\end{multicols}
\end{algorithm}

\begin{figure*}[!ht]
    \centering
    \subfloat[Role-based Transformation]{
        \includegraphics[trim=0 0 0 21,clip,width=0.5\linewidth]{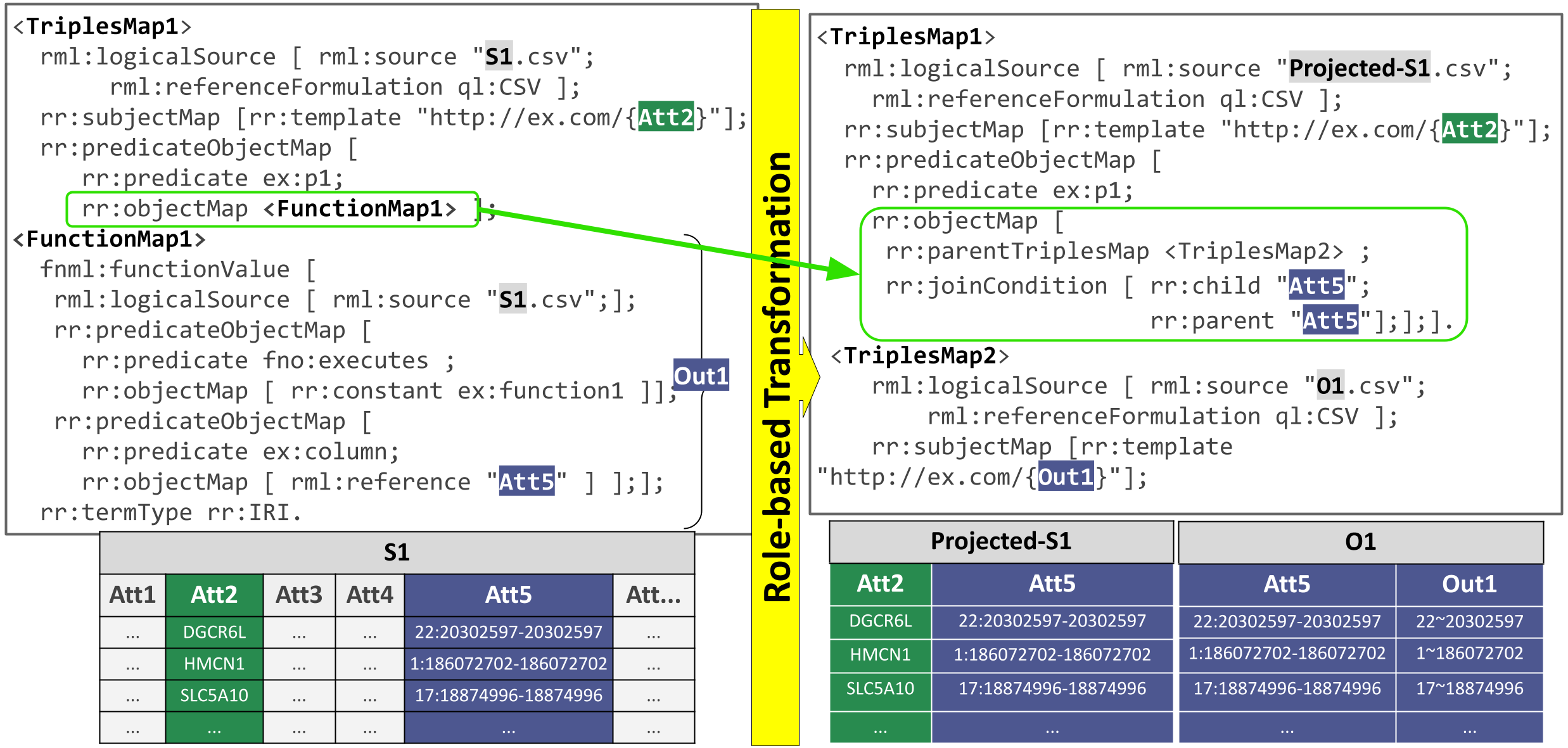}
        \label{fig:t3}
    }
    \subfloat[Attribute-based Transformation]{
        \includegraphics[trim=0 0 0 21,clip,width=0.5\linewidth]{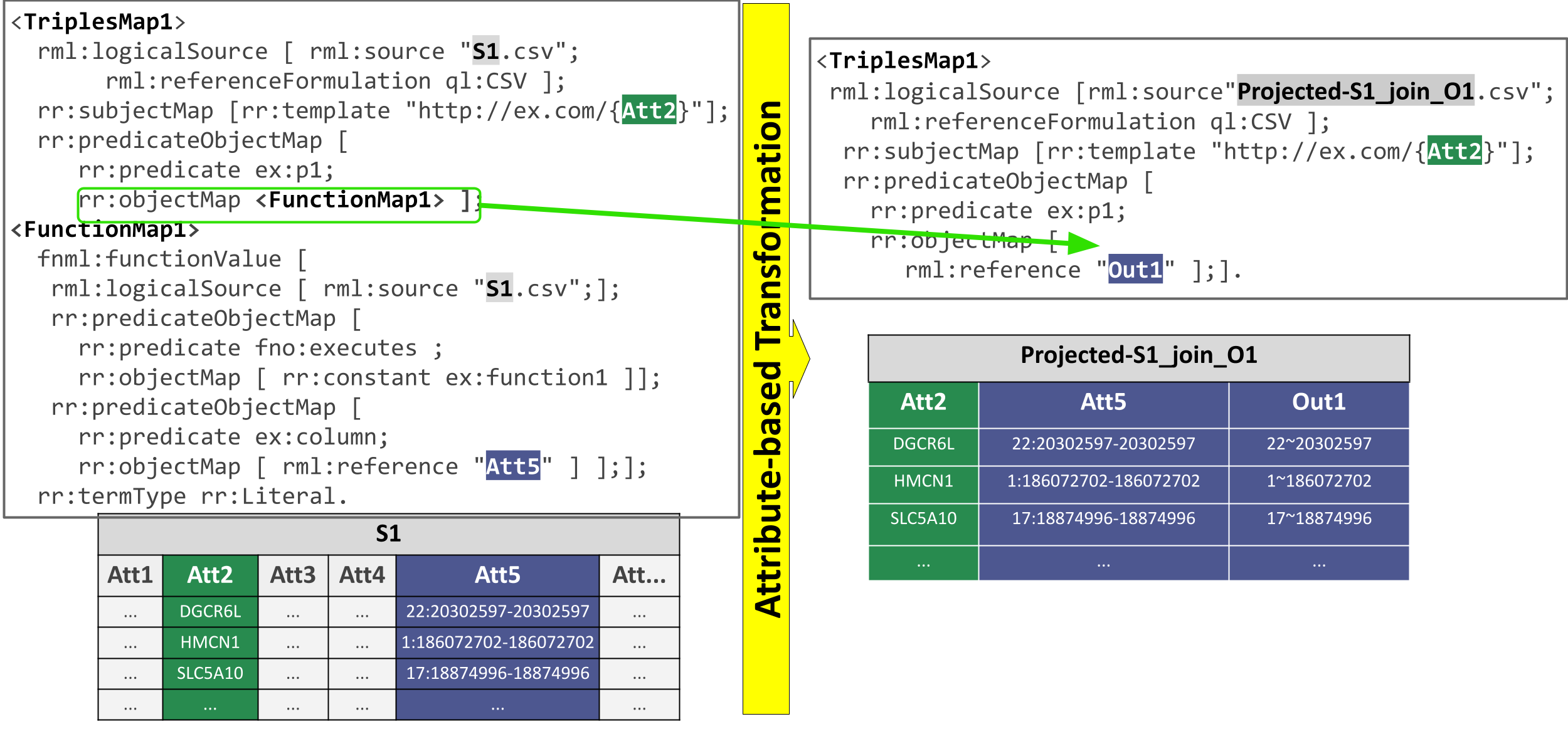}
        \label{fig:t4}
    }
    \caption{\textbf{Dragoman Role-based and Attribute-based Transformations}}
    \label{fig:transformations2}
\end{figure*}

\noindent\textbf{Role-based Transformation.} Contrary to the previous transformation, role-based transformation is performed once the output of functions (\verb|fnml:Function|-\verb|TermMap|) are utilized to build the value of a role mapping assertion (\verb|rr:objectMap|). In contrast to concept-based transformation, role-based transformation forces the join between the data sources consisting of the output of the function and the outcome of the source-based projection, to the mapping assertions. In other words, the output of the function evaluation is stored in a separate new data source and the newly generated  data source by the source-based projection. Consequently, the role mapping assertion is transformed to multi-sources role mapping assertions to replace the old data source with two data sources, i.e., the data source, including the output of the functions and the data source generated as the result of the source-based projection. Figure \autoref{fig:t3} illustrates an example of role-based transformation. As shown in the right-hand side of the figure \autoref{fig:t3}, the transformed data sources include two newly generated data sources, and the role mapping assertion is transformed to a multi-sources mapping assertion, applying \verb|joinCondition|. An outline of the implementation of the role-based transformation rule is provided in Algorithm\ref{alg:alg_role}.

\begin{algorithm}[H]
\centering
\caption{Role-based Transformation Algorithm}
\label{alg:alg_role}
\begin{multicols}{2}
\begin{algorithmic}[1]
\Require $ma,\overline{TS}$
\Ensure $Role\_Transformation(ma,\overline{TS})$
\State $NEW\_\overline{Tma}\gets\{ma\}$
\State $OLD\_\overline{Tma}\gets NEW\_\overline{Tma}$
\State Suppose $ma:$
\State $S_i(\overline{X}_{i,1}):-P(f_1(t_1),f_2(g(\overline{T}_2)))$
\State where $g$ is a \emph{term} that is a user-defined 
\State function and receives $\overline{T}_2$, a set of \emph{term}s
\State as arguments: $\overline{T}_2:\{t_{2,1},...,t_{2,m}\}$
\If {$g(.)=a(b(.))$}
\State $S_g(\overline{X}_g,b(\overline{T}_2))$ is defined for all the values of 
\State $b(\overline{T}_2)$ such that $eval(S_g(\overline{X}_g,b(\overline{T}_2)))=True$
\State $ma':S_i(\overline{X}_{i,1}),S^{Cma'}_g(\overline{X}_g,b(\overline{T}_2)),\theta_g(\overline{X}_i,1,\overline{X}_g)$
\State $:-P(f_1(t_1),f_2(t'_2))$ where $t'_2$ is a term: $t' \in S_g$
\State $Cma':\ S_g(\overline{X}_g,b(\overline{T}_2)):-C'(f_2(t'_2))$
\Else
\State $S_g(\overline{X}_g,\overline{T}_2)$ is defined for all the values of $\overline{T}_2$ 
\State such that $eval(S_g(\overline{X}_g,\overline{T}_2))=True$
\State $ma': S_i(\overline{X}_{i,1}),S^{Cma'}_g(\overline{X}_g,\overline{T}_2),$ $\theta_g(\overline{X}_i,1,\overline{X}_g)$
\State $:-P(f_1(t_1),f_2(t'_2))$ where $t'_2$ is a term: $t' \in S_g$
\State $Cma':\ S_g(\overline{X}_g,\overline{T}_2):-C'(f_2(t'_2))$
\EndIf
\State $NEW\_\overline{Tma}\gets \{ma',Cma'\}$
\State $\overline{TS}\gets \overline{TS} \cup \{S_g\}$ \\
\Return $NEW\_\overline{Tma},\overline{TS}$
\end{algorithmic}
\end{multicols}
\end{algorithm}

\noindent\textbf{Attribute-based Transformation.} This transformation is required when the output of the user-defined function is used as the \emph{term} value of an Attribute mapping assertion. Similar to the concept-based transformation, data sources are transformed by joining the data sources consisting of the output of the function and the result of the source-based projection on $S$. Consequently, the mapping assertion is transformed to include the newly generated joined data source as the replacement for the original data source. As it can be observed in the example shown in figure \autoref{fig:t4} using the output of user-defined functions - \verb|fnml:FunctionTermMap|- as the \emph{term} value of both role and attribute mapping assertions are very similar in [R2]RML.; they are differentiated by the values \verb|rr:IRI| and \verb|rr:Literal| for \verb|termType| in case of role and attribute mapping assertions, respectively. The attribute-based transformation rule can be implemented following Algorithm\ref{alg:algo_att}. 

\begin{algorithm}[H]
\centering
\caption{Attribute-based Transformation Algorithm}
\label{alg:algo_att}
\begin{multicols}{2}
\begin{algorithmic}[1]
\Require $ma,\overline{TS}$
\Ensure $Attribute\_Transformation(ma,\overline{TS})$
\State $NEW\_\overline{Tma}\gets\{ma\}$
\State $OLD\_\overline{Tma}\gets NEW\_\overline{Tma}$
\State Suppose $ma: S_1(\overline{X}_{i,1}):-A(f(t_1),g(\overline{T}_2))$ and $Cma: S_1(\overline{X}_{1,1}):-C(F_1(t_1))$ where $g(.)$ is a term that is a user-defined function and receives $\overline{T}_2$, a set of \emph{term}s as arguments: $\overline{T}_2:\{t_{2,1},...,t_{2,m}\}$
\If {$g(.)=a(b(.))$}
\State $S_g(\overline{X}_g,\overline{T}_2)$ is defined for all the values of $b(\overline{T}_2)$ 
\State such that $eval(S_g(\overline{X}_g,b(\overline{T}_2)))=True$
\State $ma': S_1(\overline{X}_{i,1}),S_g(\overline{X}_g,\overline{T}_2):-A(f(t_1),t'_2)$ 
\State where $t'_2$ is a \emph{term}: $t'_2 \in S_g$
\State $Cma': S_1(\overline{X}_{1,1}),S_g(\overline{X}_g,b(\overline{T}_2)):-C(f(t_1))$
\Else
\State $S_g(\overline{X}_g,\overline{T}_2)$ is defined for all the values of $\overline{T}_2$ 
\State such that $eval(S_g(\overline{X}_g,b(\overline{T}_2)))=True$
\State $ma': S_1(\overline{X}_{i,1}),S_g(\overline{X}_g,b(\overline{T}_2)):-A(f(t_1),t'_2)$ 
\State where $t'_2$ is a \emph{term}: $t'_2 \in S_g$
\State $Cma': S_1(\overline{X}_{1,1}),S_g(\overline{X}_g,\overline{T}_2):-C(f(t_1))$
\EndIf
\State $NEW\_\overline{Tma}\gets \{ma',Cma'\}$
\State $\overline{TS}\gets \overline{TS} \cup \{S_g\}$\\
\Return $NEW\_\overline{Tma},\overline{TS}$
\end{algorithmic}
\end{multicols}
\end{algorithm}

\begin{figure}[h]
\centering
\includegraphics[width=1.0\linewidth]{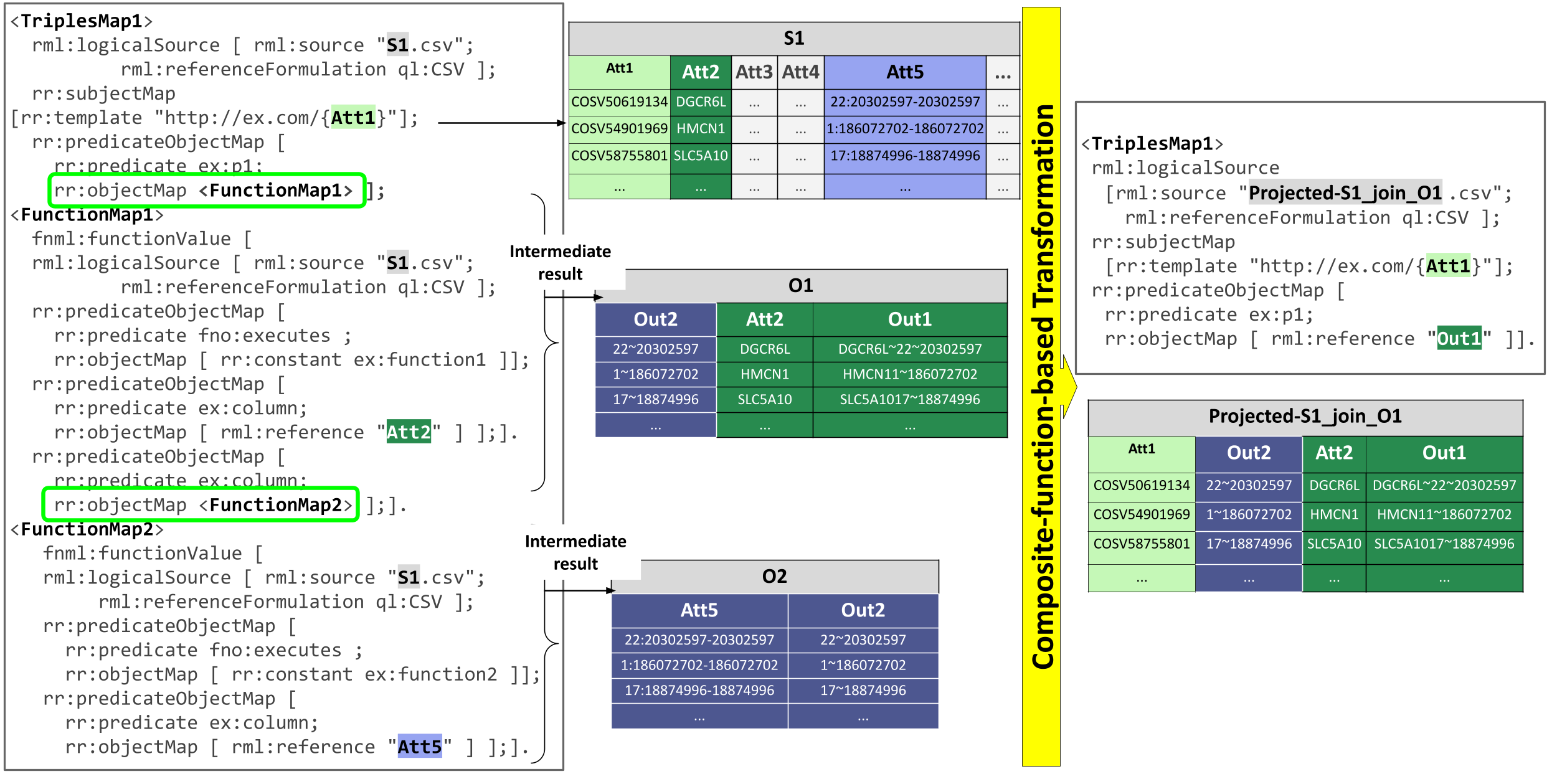}
\caption{\textbf{Composite-function-based Transformation}.}
\label{fig:t5}
\end{figure}

\noindent\textbf{Composite-Function-based Transformation.} When the user-defined function is a composite function, i.e, the output of a \verb|fnml:FunctionTermMap| is an argument to another \verb|fnml:FunctionTermMap|, the same transformation is performed independent of the type of the mapping assertion that refers to the output of the function. composite-function-based transformation is similar to the concept-based transformation; data sources are transformed by joining the data source generated by the output of all functions involved in the composite function, and the data source result from the source-based projection. Accordingly, Dragoman starts executing the composite function from the inner function, i.e., the simple function. As illustrated in the example shown at the bottom of the Figure\ref{fig:t5}, after evaluating the inner function \verb|FunctionMap2|, the materialized join between the output and the input attributes of the inner function is provided. This join data source is given to the outer function \verb|FunctionMap1| as the input. After evaluating the outer function, the composite-function-based transformation is performed along with the source-based projection.

\section{Experimental Study}
\label{sec:experiments}

\noindent We empirically assess the performance of Dragoman to answer the following questions: \begin{inparaenum}[\bf {\bf RQ}1\upshape)]
\item What is the impact of applying Dragoman in a knowledge graph creation pipeline in terms of execution time? 
\item What parameters illustrate the advantages of applying Dragoman in knowledge graph creation pipelines?
\end{inparaenum} For this purpose, we set up 220 different pipelines of knowledge graph creation, half of which deploy Dragoman to perform the tasks of functions evaluation and DIS transformation, following an eager evaluation. While the rest of the pipelines rely on the same engines which generate the RDF triples to also perform the task of function evaluations, based on a lazy evaluation strategy. With the mentioned comparison, we aim to observe whether the execution time required to generate the same knowledge graph reduces when the proposed transformations are applied in the pipeline.

\subsection{Knowledge Graph Creation Impacting Parameters and Testbeds}
\noindent We devise testbeds that include the parameters that impact the performance of different tasks in a knowledge graph creation pipeline~\cite{jozashoori2019mapsdi,chaves2019parameters,iglesias2020sdm,jozashoori2020funmap,arenas2022morph,iglesias2022scaling}. Accordingly, we consider eight parameters belonging to different elements of DIS, i.e., data sources, mapping assertions, and (user-defined) functions. We group the studied parameters into three categories. In the following, we explain the different values studied for each parameter in detail.
\subsubsection{Data Source Parameters}
\noindent\textbf{Par1: Data size.} The first parameter whose impact needs to be studied is the data source size in terms of the number of records integrated into the course of the knowledge graph creation pipeline. For this purpose, we consider data sources of 10,000, 100,000, and 1,000,000 records in CSV format. It should be noted that only several attributes in each data source may participate in the pipeline of the knowledge graph creation, i.e., being applied in mapping assertions. Accordingly, for each experiment setup, i.e., a combination of data source, mapping assertion, and function, we prepare the data source such that its total number of  attributes is 3-20 times the number of attributes utilized in mapping assertions. \textbf{Par2: Join selectivity.} The level of the selectivity of the join in multi-sources role mapping assertions is an influential parameter that is essential to be studied. For this purpose, for each data size, we prepare data sources with three levels of join selectivity: low or 80\% selectivity rate, medium or 50\%, and finally, high or 20\% selectivity rate. 

\subsubsection{Mapping Assertion Parameters}
\noindent There are several parameters regarding the mapping assertions that impact the performance of any knowledge graph creation pipeline that we explain in the following. \textbf{Par3.} The number of different mapping assertions with the same user-defined functions that receive the same arguments can affect the performance of a tool according to the evaluation strategy it relies on. We study the impact of having two, four, and six repetitions of the same function with the same input values. \textbf{Par4.} Another parameter we need to study is the type of the mapping assertion, which includes user-defined functions, i.e., if it is a concept or a role mapping assertion. \textbf{Par5.} The overall number of role mapping assertions involving no user-defined functions that coexist with a mapping assertion that includes user-defined functions can also impact the tool's overall performance. Therefore, we repeat the same experiments with one, three, and five role mapping assertions - including no user-defined functions - to observe their impact on the performance of the pipelines. \textbf{Par6.} When a group of separated multi-sources role mapping assertions forms \emph{star join}s, that can negatively impact the performance of a mapping assertion translation tool. We set up two groups of experiments; the first group includes \emph{star join}s in their original mapping assertions. The second one consists of experiments that include no mapping assertions that form \emph{star join}s, nevertheless, \emph{star join}s appear in mapping assertions after being transformed by Dragoman. The later experiments aim to investigate whether applying Dragoman in cases where the transformed mapping assertions are more complex still offers execution time savings. We study each \emph{star join} based on the three mentioned selectivity levels. \textbf{Par7.} Another complex case that can arise due to having more than one multi-sources role mapping assertion is once they form \emph{chain join}. To study the impact of this facet, we set up a multi-sources mapping assertion such that their transformed mapping assertions provided by Dragoman include \emph{chain join}.

\subsubsection{Function Parameters}
\noindent\textbf{Par8.} We assume that the type of user-defined functions can impact the performance of a mapping assertions translator. In these experiments, we consider two types of functions; non-injective surjective (NonInjSurj), and bijective. The descriptions of the examples of each type are provided in \autoref{tab:functions}. \textbf{Par9.} Another parameter that we presume to affect the execution of user-defined functions in mapping assertions is the complexity of functions, i.e., being simple or composite in terms of their arguments. Accordingly, we consider both simple and composite functions in our experimental studies.

\begin{table}[]
\scriptsize
\centering
\caption{Examples of bijective and non-injective surjective user-defined functions.}
\label{tab:functions}
\resizebox{1.0\linewidth}{!}{
\begin{tabular}{ |c|c|c|c|}
\rowcolor[HTML]{4D5790} 
\hline
 \rowcolor[HTML]{4D5790} 
\textcolor[rgb]{1.0,1.0,1.0}{\textbf{Function Type}} & \textcolor[rgb]{1.0,1.0,1.0}{\textbf{Function Name}} & \textcolor[rgb]{1.0,1.0,1.0}{\textbf{Pre-Condition}} & \textcolor[rgb]{1.0,1.0,1.0}{\textbf{Post-Condition}}  \\
\hline
\cellcolor[HTML]{CACBFF} \textbf{Bijective} &  \cellcolor[HTML]{CACBFF} \textbf{reverseString()} &
\cellcolor[HTML]{CACBFF} A case-insensitive string &
\cellcolor[HTML]{CACBFF} A case-insensitive string that is the exact reverse of the input string \\
\cellcolor[HTML]{CACBFF} \textbf{Non-Injective Surjective} & 
\cellcolor[HTML]{CACBFF} \textbf{toLower()} &
\cellcolor[HTML]{CACBFF} A case-sensitive string &
\cellcolor[HTML]{CACBFF} The exact string as the input string in in lower cases \\
\hline

\end{tabular}
}
\end{table}

\subsection{Implementation}
\noindent \textbf{Dragoman} is implemented in Python3. As a proof of concept, the implementation includes the transformation of two data source formats: CSV files and relational databases (RDB). It should be noted that there are small differences in implementing the transformation rules between CSV and RDB. As explained previously, the data source resulting from the concept-based transformation is the join between two data sources. In the case of having data sources as CSV files, Dragoman stores the materialization of the joins between the sources in CSV files. However, in the case of having the data sources as relational tables, instead of materializing the joins, we add the SQL join queries in the mapping assertions enabling real-time materialization. Dragoman is open-source and licensed under Apache License 2.0. It is publicly accessible through a GitHub repository\furl{https://github.com/SDM-TIB/Dragoman} and Zenodo\furl{https://doi.org/10.5281/zenodo.6418124}.\\ 

\noindent \textbf{Functions.} As we explained earlier, the same experiments are repeated twice to compare the execution time of different knowledge graph creation pipelines in the presence and absence of Dragoman. In other words, we create the same knowledge graph, once using Dragoman to perform the evaluation of the functions and then another engine to generate the RDF triples, and the second time, only using an engine to perform both tasks of the function evaluation and RDF triples generation. In that regard, we can only use RML-compliant engines that are capable of executing user-defined functions. We also consider the empirical study conducted in ~\cite{iglesias2020sdm} and ~\cite{jozashoori2020funmap} to only select the engines that are efficient in executing multi-sources role mapping assertions, i.e., SDM-RDFizer\furl{https://github.com/SDM-TIB/SDM-RDFizer} and RocketRML\furl{https://github.com/semantifyit/RocketRML} \footnote{RMLMapper is another engine capable of executing functions, while, it is inefficient executing multi-sources role mapping assertions according to the experimental results reported in ~\cite{iglesias2020sdm,jozashoori2020funmap}}. Furthermore, the implementation of test functions is required to be added to the chosen engines in order to equip them with the evaluation of these functions.  Following the languages in which these engines are developed, we implement our test functions in SDM-RDFizer using Python and in RocketRML using Javascript.

\begin{table*}[h!]
\centering
\caption{}
\label{tab:experiments}
\resizebox{1.0\linewidth}{!}{
\begin{tabular}{|cccccc|}
\hline
 & & & \multicolumn{3}{c|}{Experiment Groups}\\
\hhline{~~~---}
\multicolumn{3}{|c}{Studied Parameters} & Efficiency & Complex & Composite\\
 & & & & Joins & Functions\\
\hline
 & & 10K & \xmark & \xmark & \xmark \\
\hhline{~~----}
 & Par1: Data Size & 100K & \xmark & \xmark & \xmark \\
\hhline{~~----}
Data & & 1M & \xmark & \xmark & \xmark \\
\hhline{~-----}
Source & & Low & & \xmark & \xmark \\
\hhline{~~----}
 & Par2: Selectivity & Medium & & \xmark & \xmark \\
\hhline{~~----}
 & & High & & \xmark & \xmark \\
\hline
 & & 2 & & \xmark & \\
\hhline{~~----}
 & Par3: \# of appearances of the same & 3 & \xmark & & \\
\hhline{~~----} 
 & user-defined functions & 4 & & \xmark & \\
\hhline{~~----}
 & & 5 & \xmark & & \\
\hhline{~-----}
 & Par4: Type of mapping assertions that & Concept Mapping Assertion & & \xmark & \xmark\\
\hhline{~~----}
Mapping & include user-defined functions & Role Mapping Assertion & \xmark & \xmark & \\
\hhline{~-----}
Assertion & & 1 & \xmark & & \\
\hhline{~~----}
 & Par5: The overall \# of role mapping assertions & 3 & \xmark & & \\
\hhline{~~----}
 & & 5 & \xmark & & \\
\hhline{~-----}
 & Par6: Star Join & Four multi-sources role mapping & \xmark & \xmark & \\
\hhline{~~~~~~}
 & & assertions with the same $MJ$ &  &  & \\
\hhline{~-----}
 & Par7: Chain Join & 1 & & \xmark & \\
\hhline{------}
 & Par8: Types of user-defined functions & Non-Injective Surjective & \xmark & & \\
\hhline{~~----}
Function & & Bijective & \xmark & & \\
\hhline{~-----}
 & Par9: Composite user-defined functions & a(b(c(.)))
 & & & \xmark \\
\hline
\end{tabular}
}
\end{table*}

\subsection{Experimental Setups}
\noindent Considering the number of parameters required to be studied, the number of control variables for each experiment is also significant. To better observe and understand, we categorize the experiments into three groups according to the three most challenging parameters that have not been considered in the similar studies~\cite{jozashoori2020funmap} i.e., function type, composite function, and complex multi-sources role mapping assertions.\\

\noindent \textbf{Setups.} To answer RQ1, we define the process of knowledge graph creation as the combination of two tasks, including the execution of the functions and the translation of the mapping rules into RDF triples. Therefore, we execute the knowledge graph creation pipeline twice for each experiment setup. The first execution considers the same RML-compliant engine to perform both the tasks of the function execution and RDF triples generation. However, in the second attempt, the tasks of executing the functions and providing function-free mapping assertions are assigned to Dragoman, while the generation of the RDF triples is performed by an RML-compliant engine according to the transferred DIS provided by Dragoman.\\

\begin{figure*}[t!]
    \centering
    \subfloat[10k records - NonInjSurj]{
        \includegraphics[trim=0 0 0 21,clip,width=0.33\linewidth]{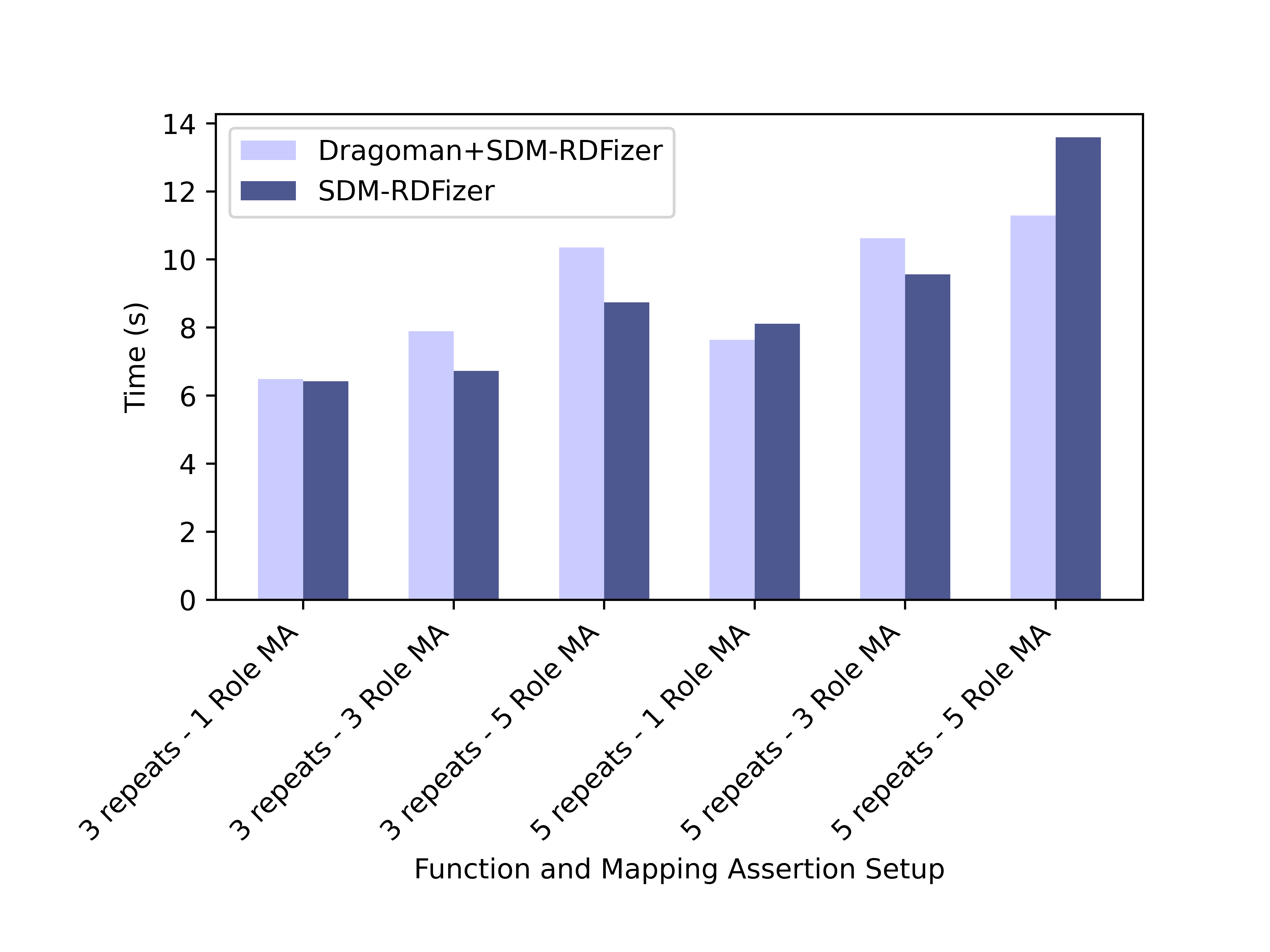}
        \label{fig:10k_nonInj_SDM-RDFizer}
    }
    \subfloat[100k records - NonInjSurj]{
        \includegraphics[trim=0 0 0 21,clip,width=0.33\linewidth]{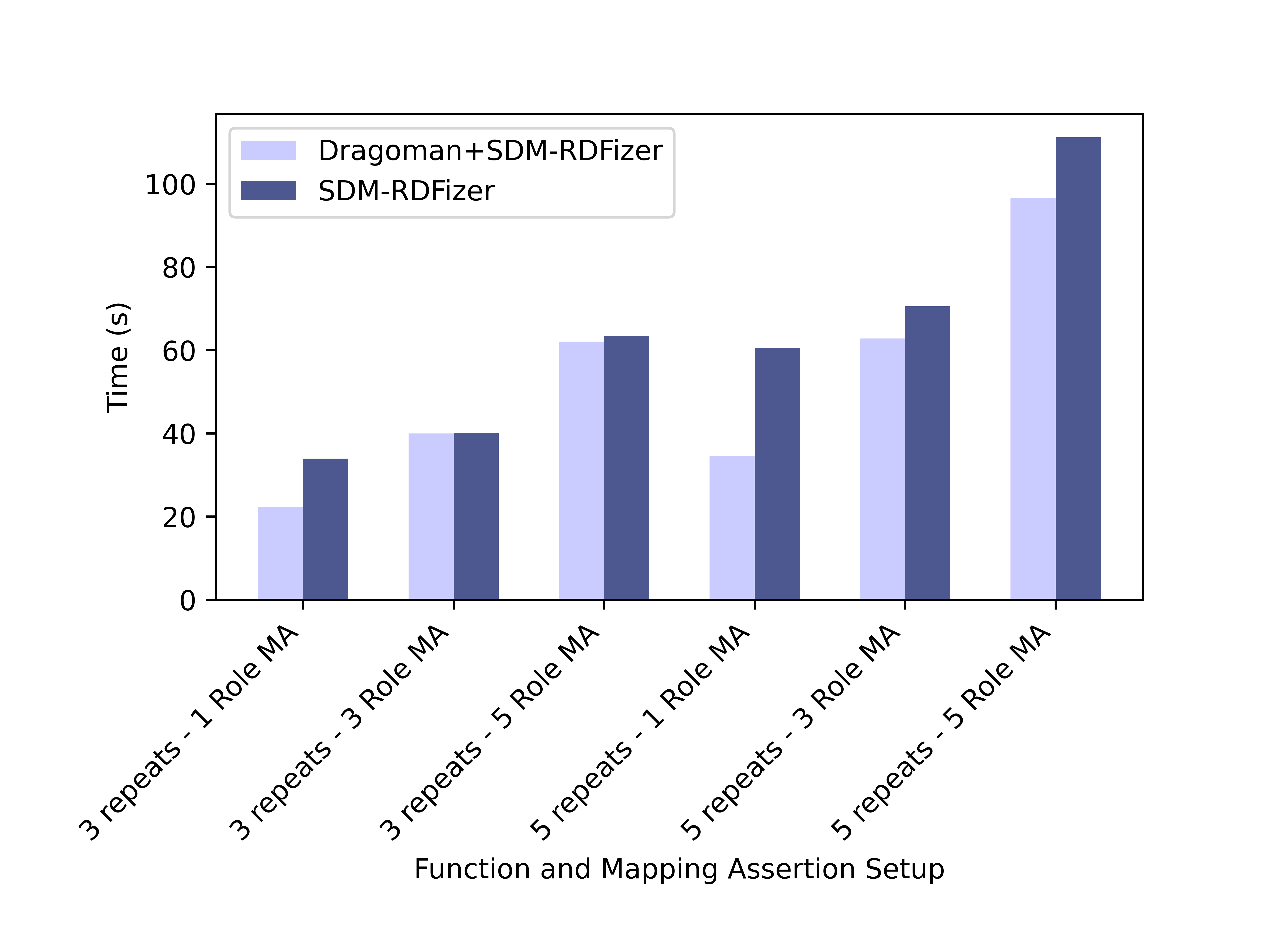}
        \label{fig:100k_nonInj_surj_SDM-RDFizer}
    }
    \subfloat[1M records - NonInjSurj]{
        \includegraphics[trim=0 0 0 21,clip,width=0.33\linewidth]{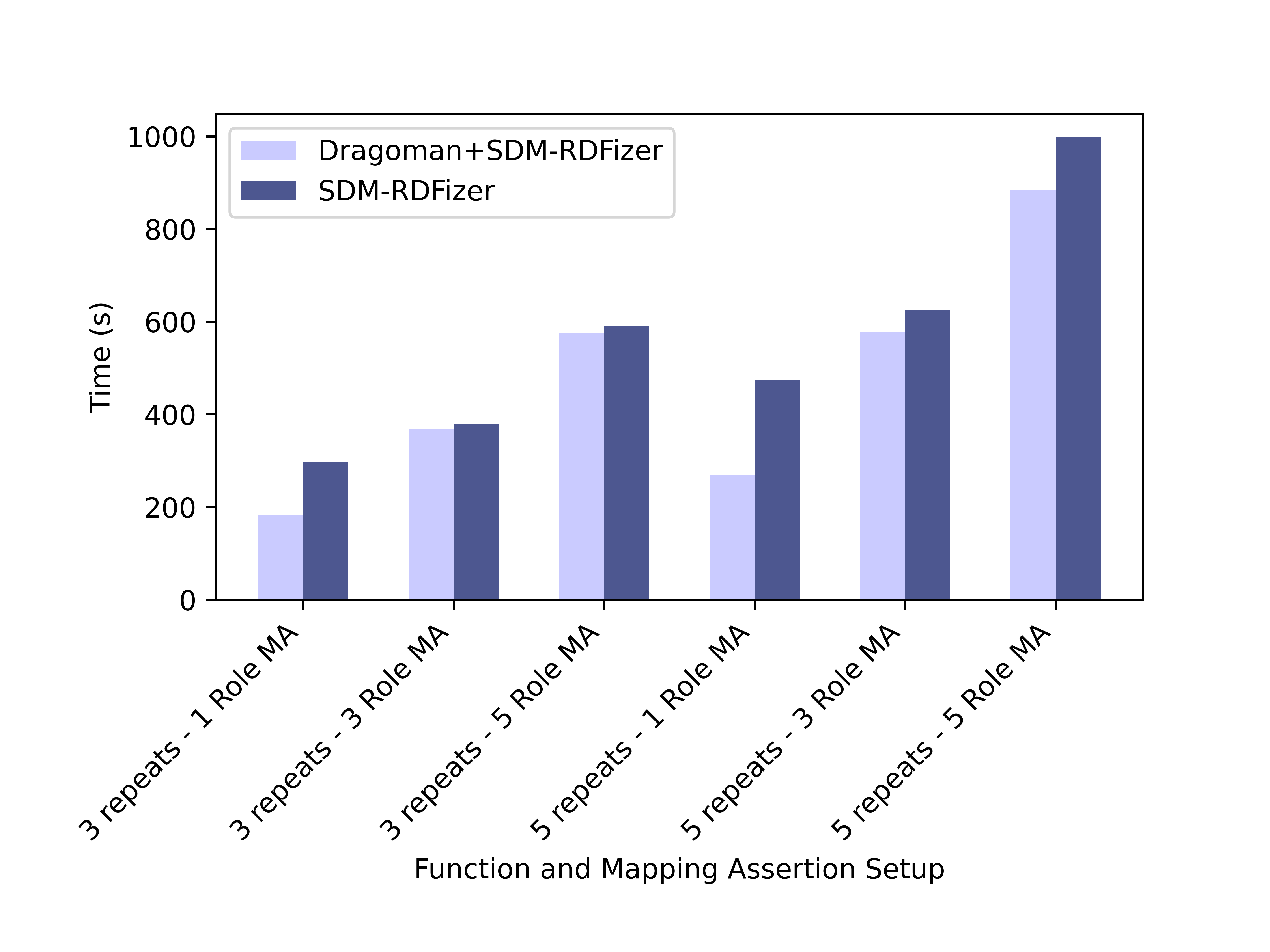}
        \label{fig:1M_nonInj_surj_SDM-RDFizer}
    }
    ~\\\vspace*{.75em}
    \subfloat[10k records - NonInjSurj]{
        \includegraphics[trim=0 0 0 21,clip,width=0.33\linewidth]{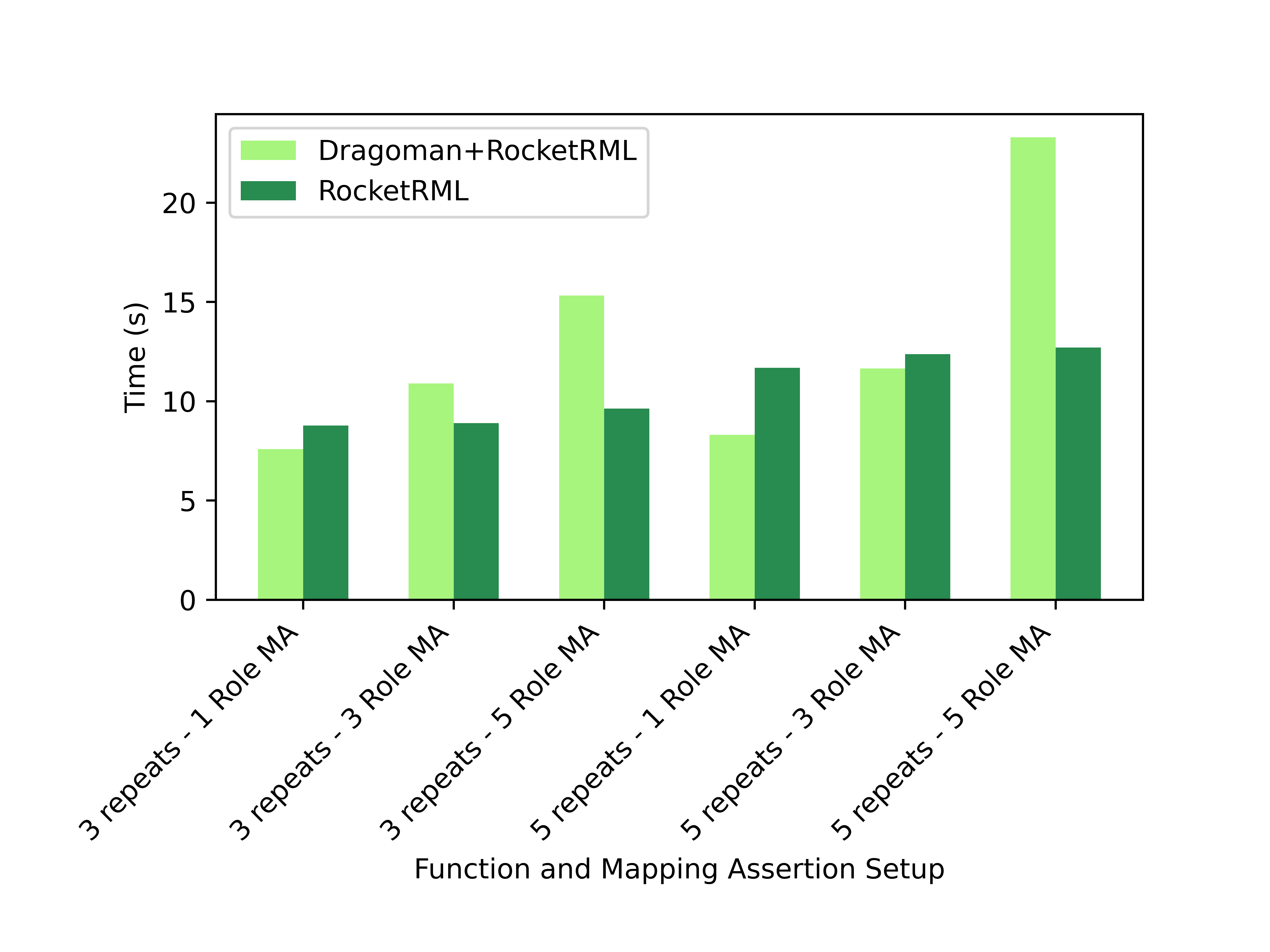}
        \label{fig:10k_nonInj_surj_Rocketrml}
    }
    \subfloat[100k records - NonInjSurj]{
        \includegraphics[trim=0 0 0 21,clip,width=0.33\linewidth]{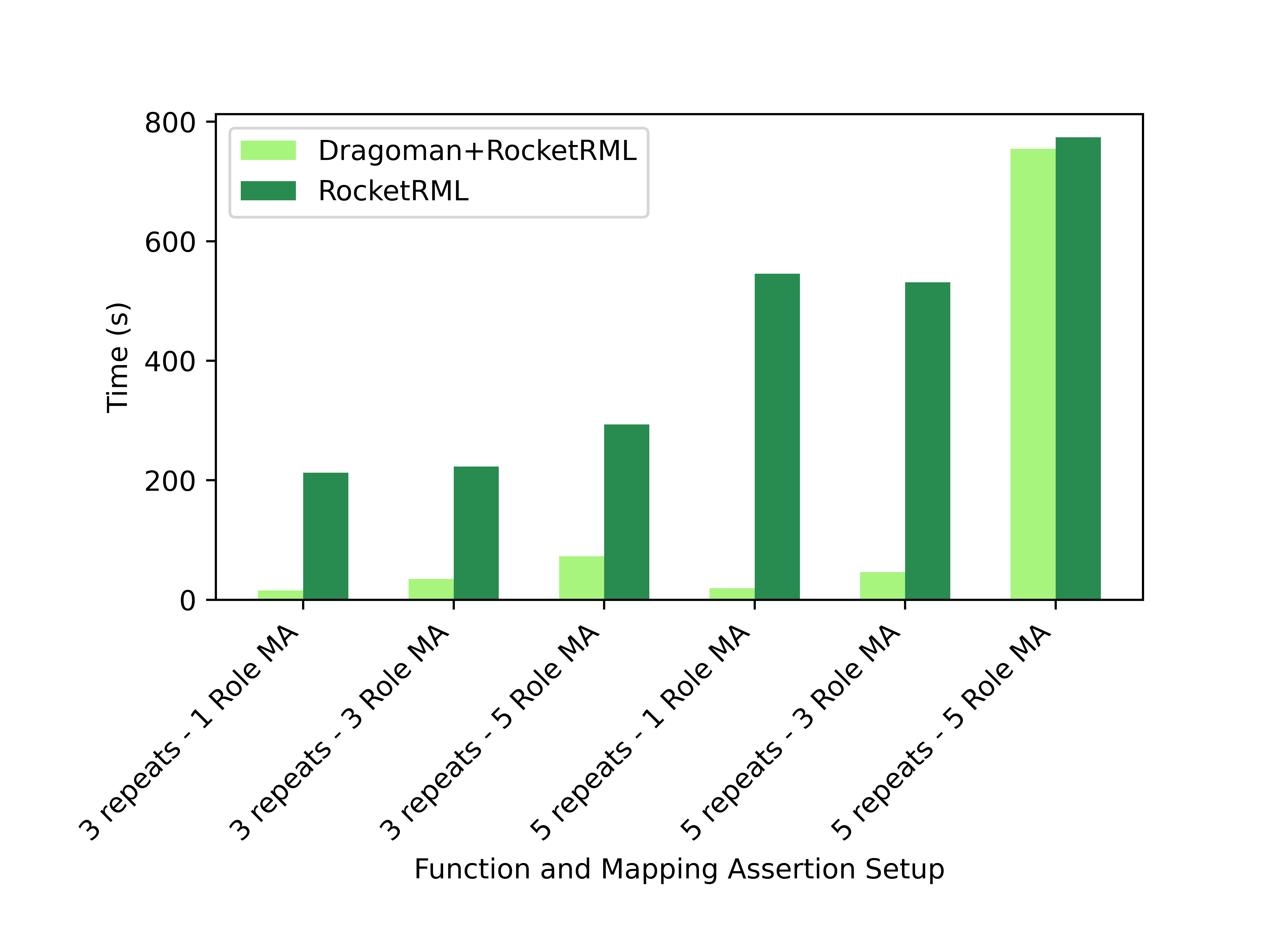}
        \label{fig:100k_nonInj_surj_Rocketrml}
    }
    \subfloat[1M records - NonInjSurj]{
        \includegraphics[trim=0 0 0 21,clip,width=0.33\linewidth]{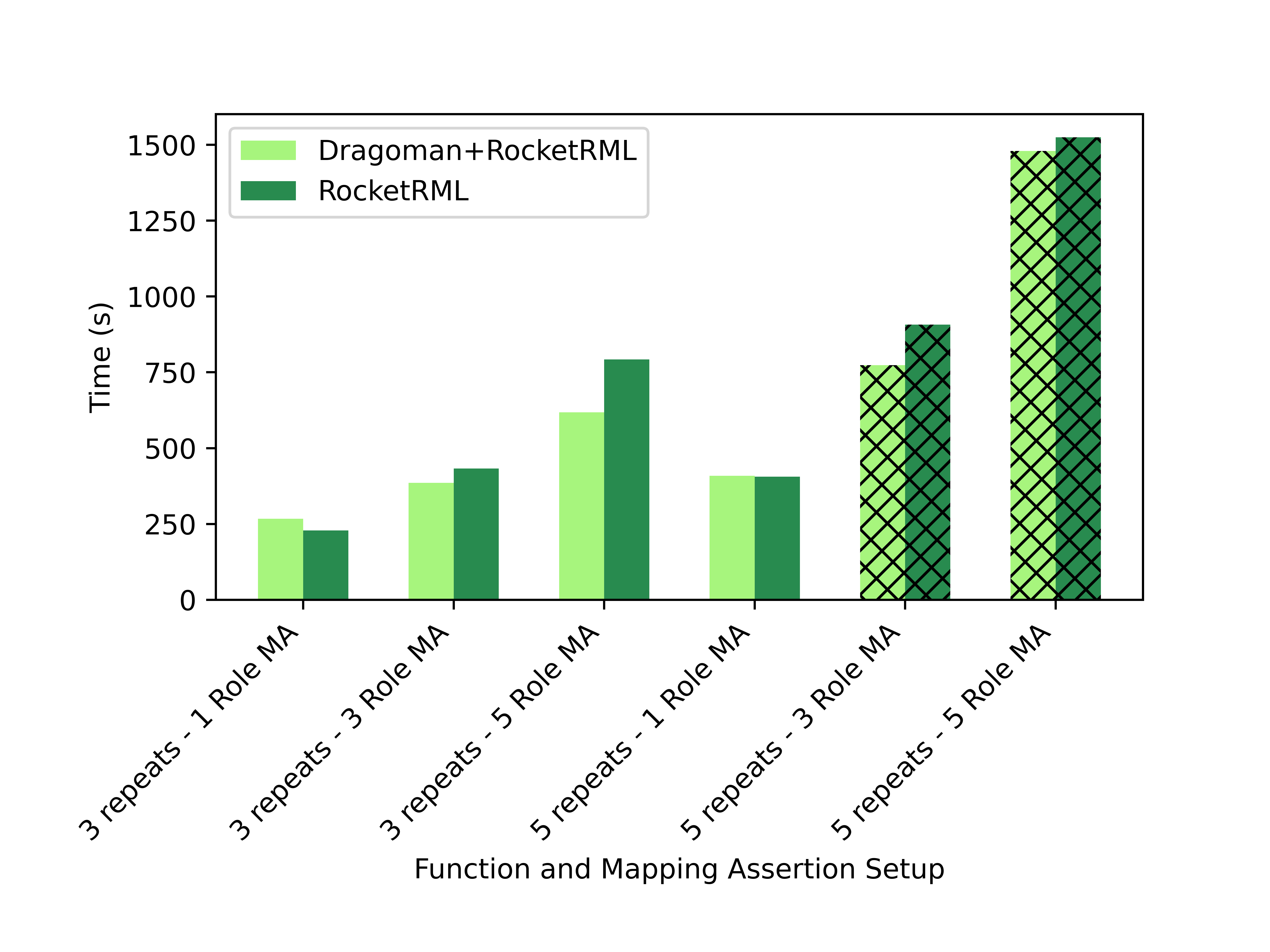}
        \label{fig:1M_nonInj_surj_Rocketrml}
    }
    \caption{Comparing the total execution time required for knowledge graph creation in presence of non-injective subjective functions, NonInjSurj.}
    \label{fig:nonInjSurj}
\end{figure*}

\noindent\textbf{Metrics.} \textit{Execution time:} Elapsed time spent by the whole pipeline, including function execution to complete the creation of a Knowledge graph; it is measured as the absolute wall-clock system time as reported by the \verb|time| command of the Linux operating system. The timeout is set to be five hours. The experiments were run in an Intel(R) Xeon(R) equipped with a CPU E5-2603 v3 @ 1.60GHz 20 cores, 64 G.B. memory, and the O.S. Ubuntu 16.04LTS.\\

\noindent\textbf{Datasets and Mapping Assertions.} To be able to create experiments considering all the parameters explained earlier, we extend the
SDM-Genomic-Datasets\furl{https://doi.org/10.6084/m9.figshare.14838342} generating a new subset of the COSMIC mutation dataset\furl{https://cancer.sanger.ac.uk/cosmic GRCh37, version90, released August 2019}. This dataset is created by including 38 attributes of the original COSMIC mutation dataset. Another dataset is created by randomly selecting records from the UMLS database composed of two attributes; label and CUI identifiers. We combine these two datasets and consider them the base dataset for our experiments. The objective of combining instances of two different data sources is to avoid any possible bias that may have been generated by a particular source in the data values that are the input of the functions. We create three datasets by randomly selecting 10k, 100k, and 1 million records from the base datasets. We create 17 sets of mapping assertions combining different types and numbers of mapping assertions and user-defined functions considering explained parameters.\\

\begin{figure}[]
    \centering
    \subfloat[10k records - Bijective]{
        \includegraphics[trim=0 0 0 21,clip,width=0.33\linewidth]{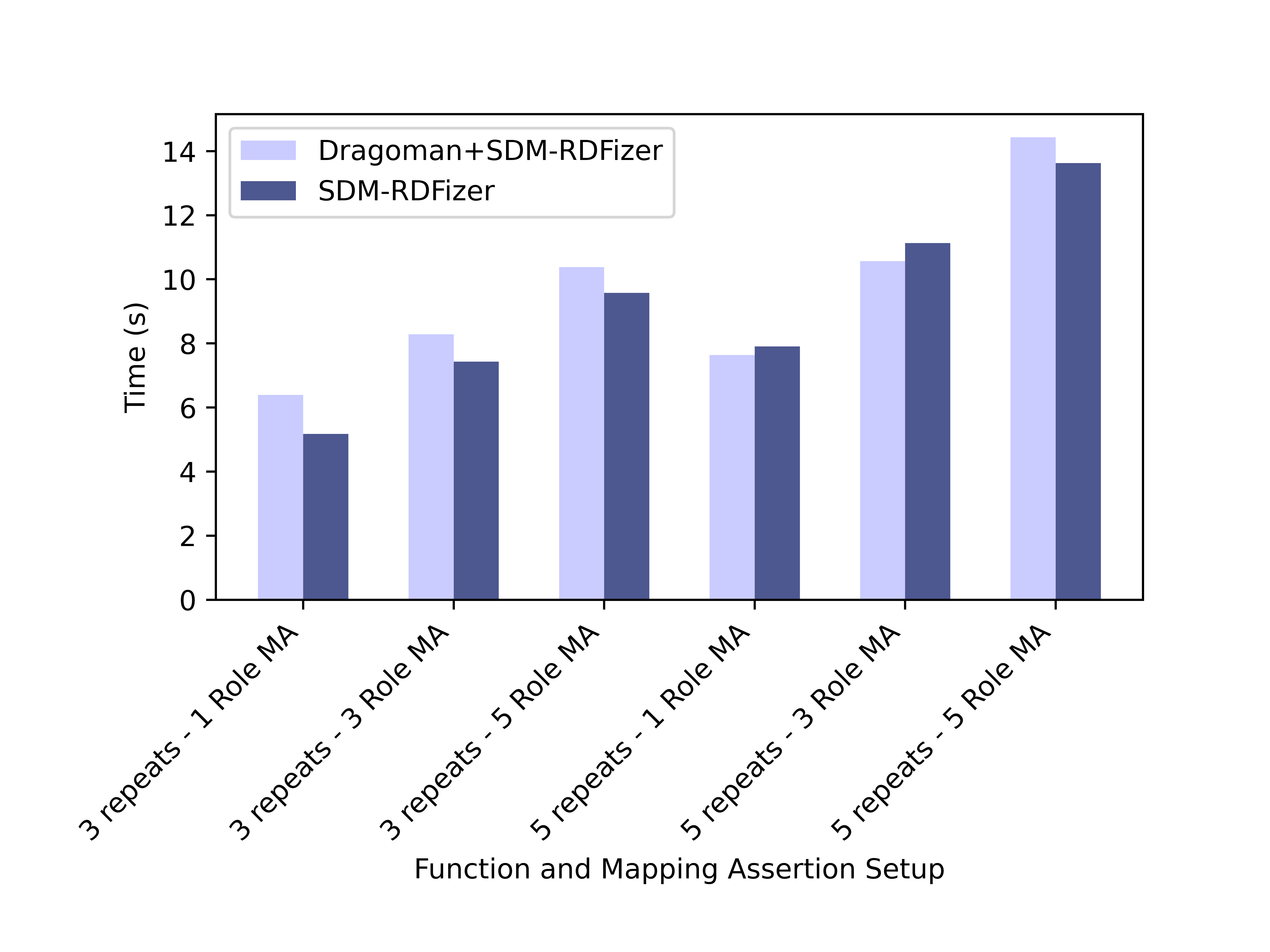}
        \label{fig:10k_bij_SDM-RDFizer}
    }
    \subfloat[100k records - Bijective]{
        \includegraphics[trim=0 0 0 21,clip,width=0.33\linewidth]{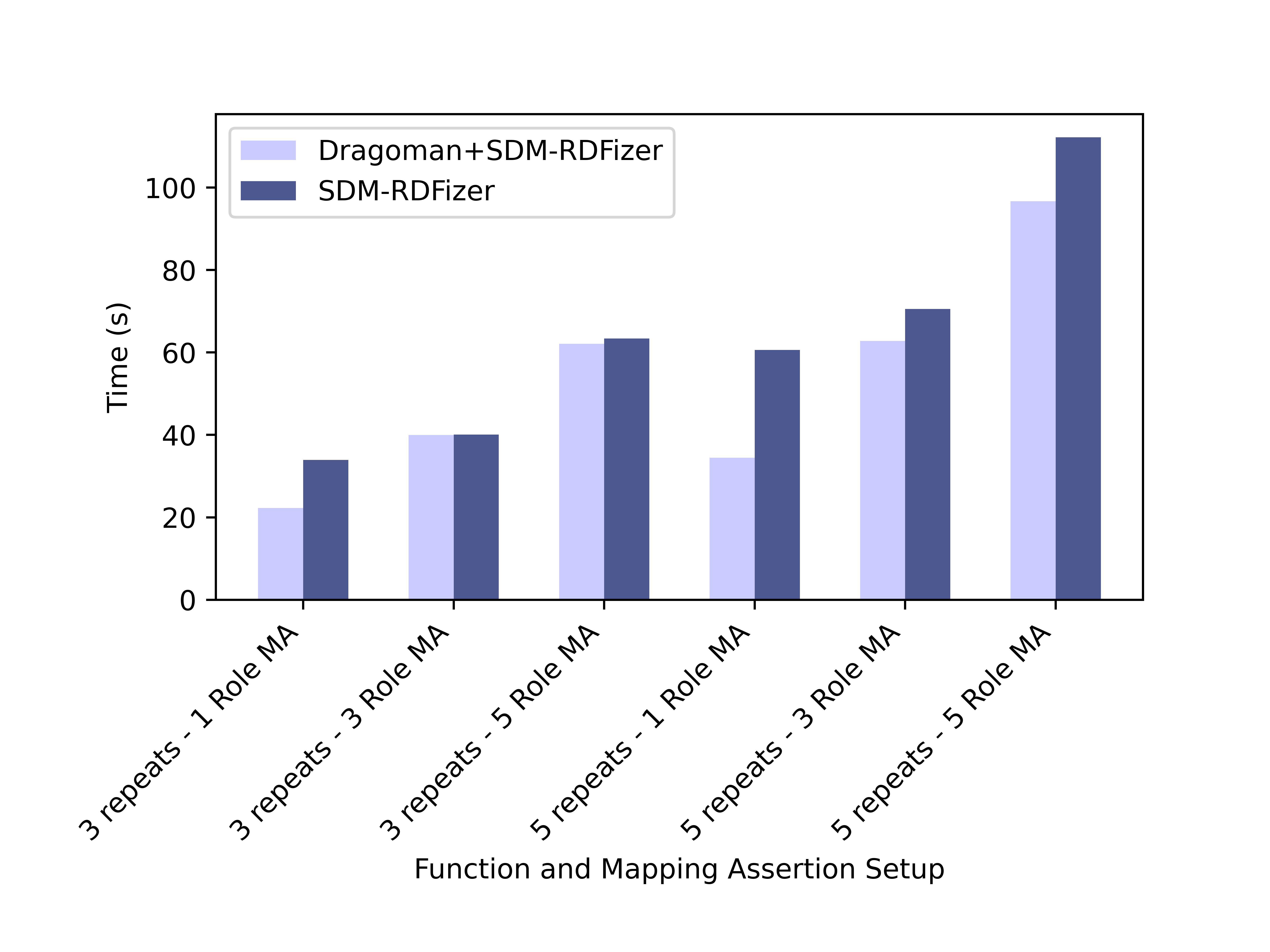}
        \label{fig:100k_bij_SDM-RDFizer}
    }
    \subfloat[1M records - Bijective]{
        \includegraphics[trim=0 0 0 21,clip,width=0.33\linewidth]{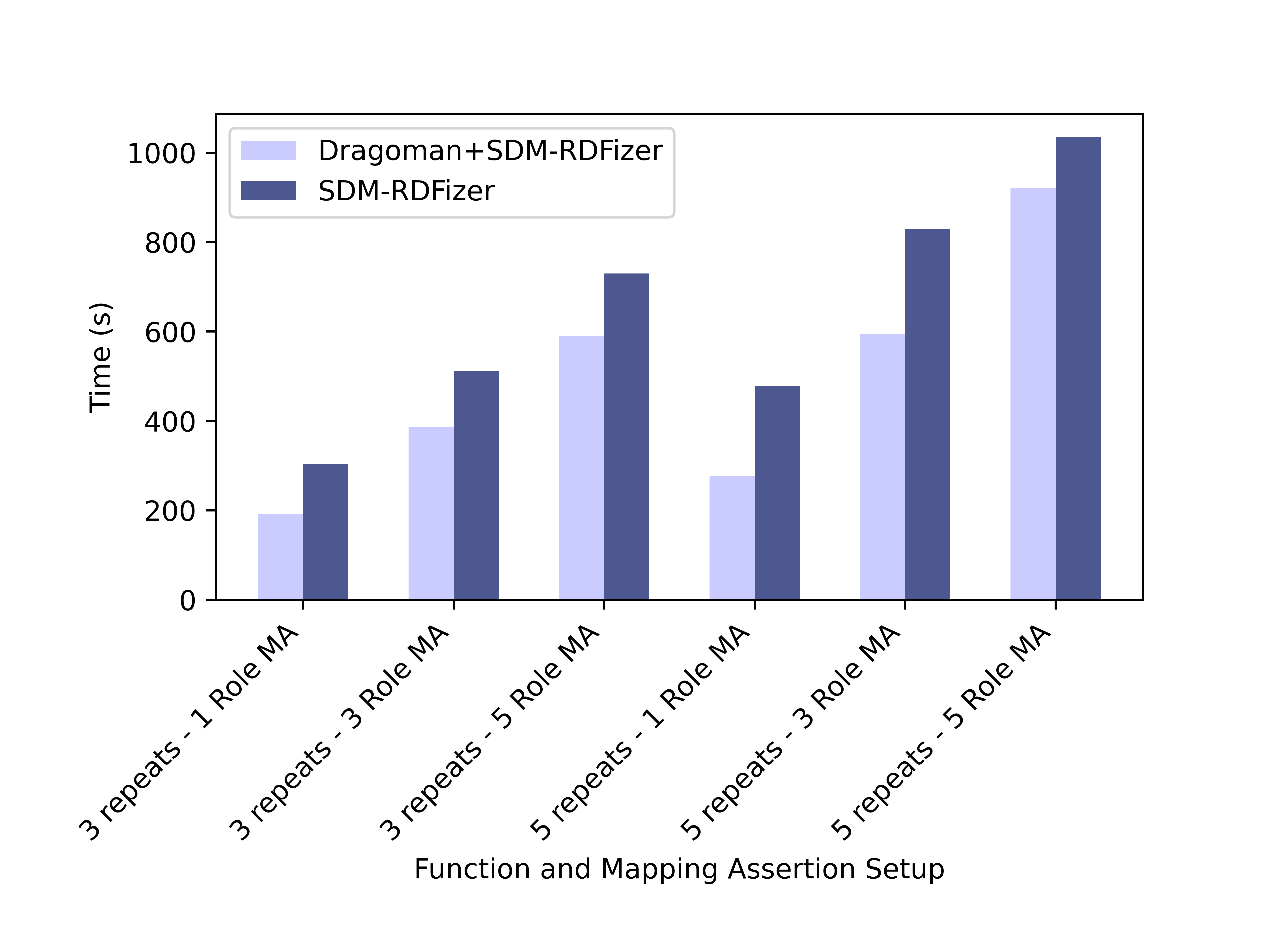}
        \label{fig:1M_bij_SDM-RDFizer}
    }
    ~\\\vspace*{.75em}
    \subfloat[10k records - Bijective]{
        \includegraphics[trim=0 0 0 21,clip,width=0.33\linewidth]{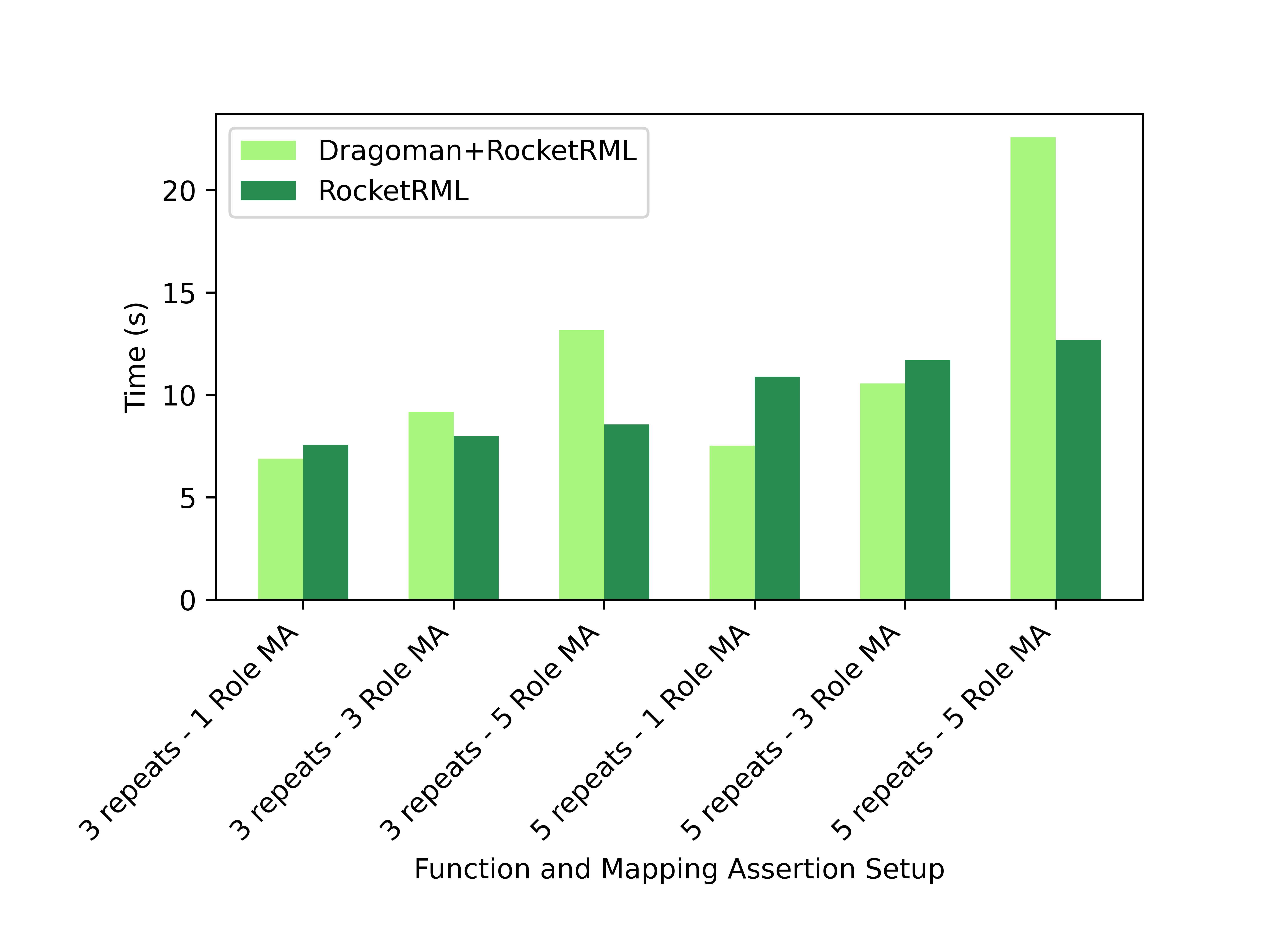}
        \label{fig:10k_bij_Rocketrml}
    }
    \subfloat[100k records - Bijective]{
        \includegraphics[trim=0 0 0 21,clip,width=0.33\linewidth]{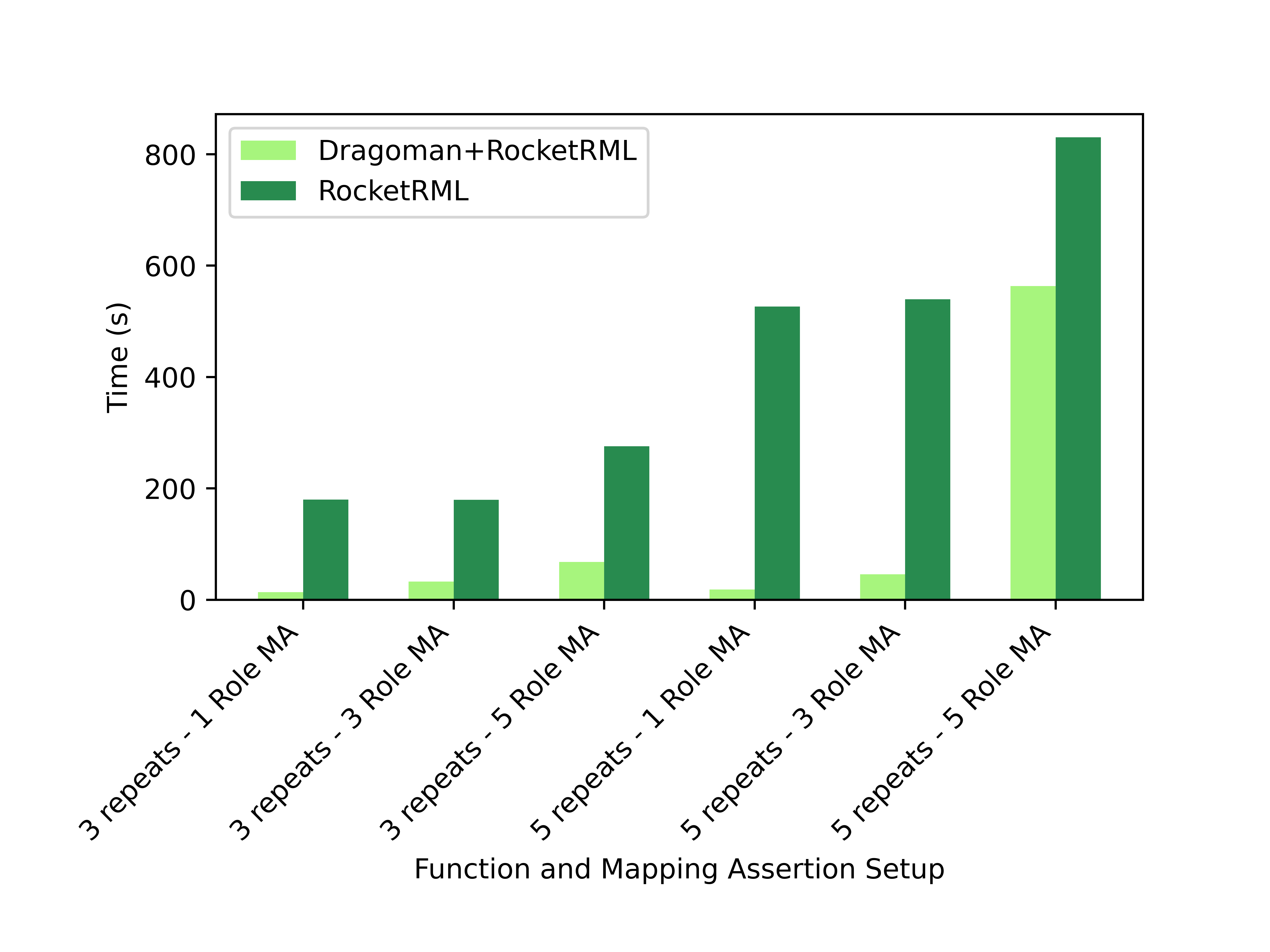}
        \label{fig:100k_bij_Rocketrml}
    }
    \subfloat[1M records - Bijective]{
        \includegraphics[trim=0 0 0 21,clip,width=0.33\linewidth]{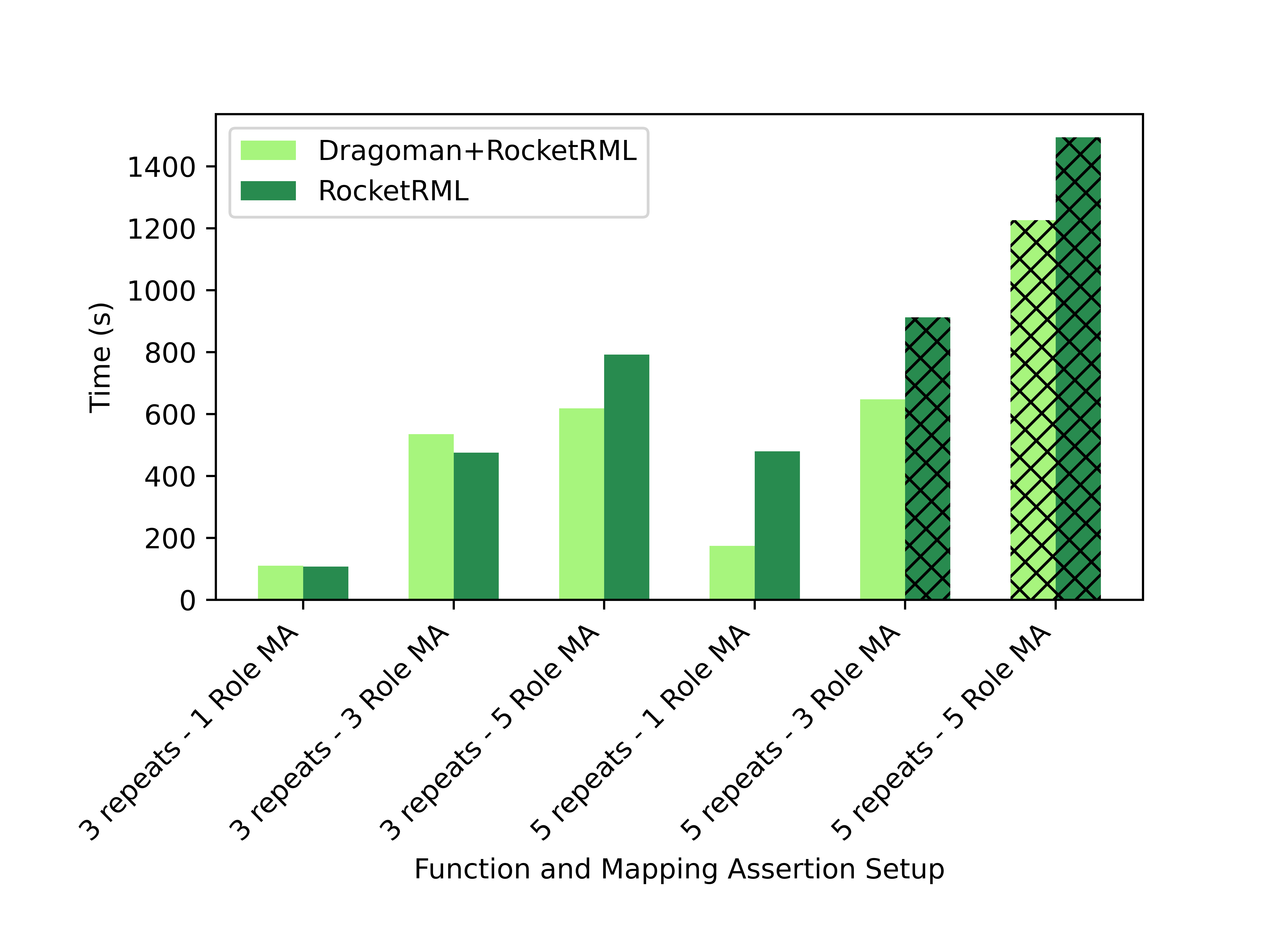}
        \label{fig:1M_bij_Rocketrml}
    }
    \caption{Comparing the total execution time required for knowledge graph creation in presence of bijective functions.}
    \label{fig:bijective}
\end{figure}

\subsubsection{Efficiency}
\label{sec:efficiency}
\noindent \textbf{Testbeds.} We set up 144 experiments in this category to study four different parameters, including Par1, Par3, Par5, and Par8. To avoid any possible bias generated by the engines translating mapping assertion engines, we evaluate the same experiments in this category with two recognized RML-compliant engines that support functions execution, i.e., RocketRML v1.12.0\furl{https://github.com/semantifyit/RocketRML} and SDM-RDFizer v4.0\furl{https://github.com/SDM-TIB/SDM-RDFizer}. In other words, we set up 72 different DIS and observed the creation of knowledge graphs from each DIS twice, using one of the two engines. It should be noted that in choosing the RML-compliant engines that support function execution, our criteria is the performance of the engines translating multi-sources role mapping assertions~\cite{iglesias2020sdm}; the two selected engines have competitive performance translating multi-sources role mapping assertions.\\

\noindent \textbf{Observations and Results.} The results of this group of experiments using the non-injective surjective function are visualized in \autoref{fig:nonInjSurj} while the results of the same experiments using the bijective function type are summarized in \autoref{fig:bijective}. \autoref{fig:10k_nonInj_SDM-RDFizer}, \autoref{fig:100k_nonInj_surj_SDM-RDFizer}, \autoref{fig:1M_nonInj_surj_SDM-RDFizer}, \autoref{fig:10k_bij_SDM-RDFizer}, \autoref{fig:100k_bij_SDM-RDFizer}, and \autoref{fig:1M_bij_SDM-RDFizer} illustrate - in purple - the results of executing the set-up knowledge graph pipelines using SDM-RDFizer as the RML-compliant, while, the other six subfigures including \autoref{fig:10k_nonInj_surj_Rocketrml}, \autoref{fig:100k_nonInj_surj_Rocketrml}, \autoref{fig:1M_nonInj_surj_Rocketrml}, \autoref{fig:10k_bij_Rocketrml}, \autoref{fig:100k_bij_Rocketrml}, and \autoref{fig:1M_bij_Rocketrml} present - in green - the results of executing the same pipelines applying RocketRML as the RML-compliant engine. The lighter colored bars in all the sub-figures \autoref{fig:nonInjSurj} and \autoref{fig:bijective} show the results utilizing Dragoman for executing the functions and transformation of the pipelines and applying the RML-compliant engine for translating the output of Dragoman into the RDF knowledge graph. In contrast, the darker colored bars in \autoref{fig:nonInjSurj} and \autoref{fig:bijective} represent the results of executing the whole knowledge graph creation pipeline, i.e., including the functions evaluations, using the RML-compliant engines. The textured bars present the results of the experiments in which the engine is unable to generate the complete result.\\ 

\noindent \textbf{Observations on Par1:} Comparing the execution time required by the pipelines, including Dragoman, with the same ones without applying Dragoman clearly shows the advantages of utilizing Dragoman in the case of large-size data sources. In the case of having small data sources, i.e., 10k, we observe no cost savings. This means that the transformations performed by Dragoman add more overheads rather than reducing the cost. However, in the case of 100k and 1M, we can observe significant benefits in applying Dragoman. It should be noted that the results of the same experiments differ when different RML-compliant engines are applied. This can be due to the different algorithms and performances that engines have, which is out of the scope of this empirical study.\\ 

\noindent \textbf{Observations on Par3 and Par5:} Comparing the performances of the pipelines with increasing numbers of the appearances of the same functions or numbers of role mapping assertions suggests that applying Dragoman can improve the performance significantly while the numbers increase. Nevertheless, applying Dragoman in the pipelines with small numbers of Par3 or Par5 can increase the required execution time.\\

\noindent \textbf{Observations of Par8:} In overall, we can observe in \autoref{fig:nonInjSurj} and \autoref{fig:bijective} that the results of the same experiments, i.e., the pipelines with the same values for the Par1, Par3, and Par5, show the same patterns using different values for Par8. In other words, these results do not support our hypothesis about the impact of the type of functions on knowledge graph creation.\\

\subsubsection{Complex Joins}
\label{sec:complex_join}
\noindent \textbf{Testbeds.} The main focus of this category of experiments is to study the impact of Par6 and Par7 in different experimental scenarios. Accordingly, these experiments can be divided into two sub-categories; Chain Join studies the impact of Par7, and Star Join focuses on Par6. Both categories study the impact of Par1 considering three different data sizes, i.e., 10k, 100k, and 1 million records. However, they both consider only one value for Par3, which is four appearances of the same user-defined function. \textit{Chain Join.} This group of experiments involves one chain join as the value of par7, however, they adopt different values of Par2, i.e., low or 80\%, medium or 50\%, and finally, high or 20\% join selectivity. This setup aims to study the impact of chain join in the presence of different selectivity rates. \textit{Star Join.} In these experiments, we focus on the Par6 parameter, a.k.a star join. Since star join is an expensive logical operation for an engine, it is very important to be studied in the empirical evaluation of any knowledge graph creation engine. According to the experimental study by Iglesias et al.~\cite{abs-2201-09694}, SDM-RDFizer can perform star joins, which RocketRML fails to do. Therefore, we utilize SDM-RDFizer in this group of experiments.\\

\begin{figure*}[t!]
    \centering
    \subfloat[Join Chain]{
        \includegraphics[trim=0 0 0 21,clip,width=0.6\linewidth]{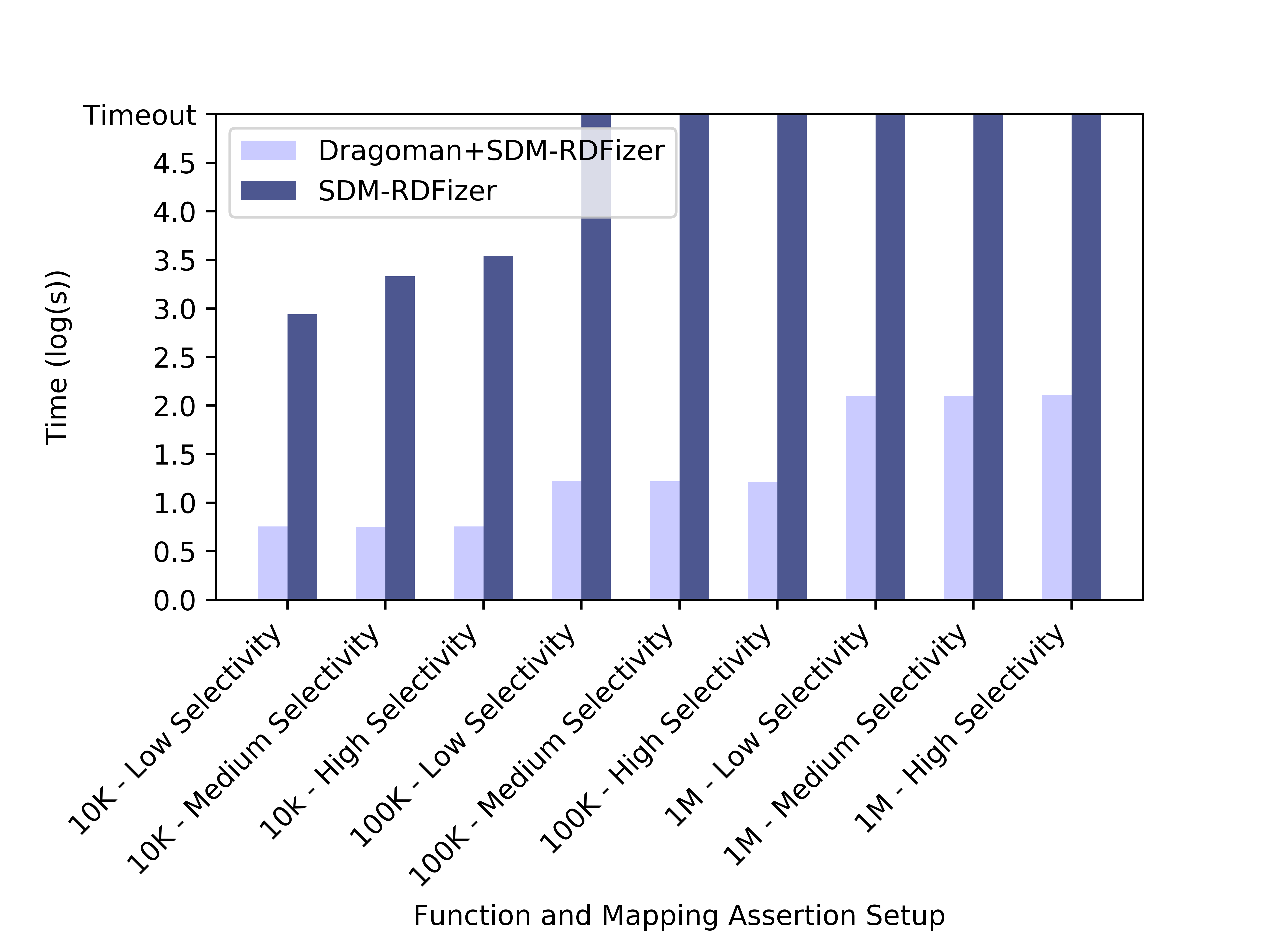}
        \label{fig:chainJoin}
    }
        ~\\\vspace*{.75em}
    \subfloat[Star Join - Simple Function]{
        \includegraphics[trim=0 0 0 21,clip,width=0.5\linewidth]{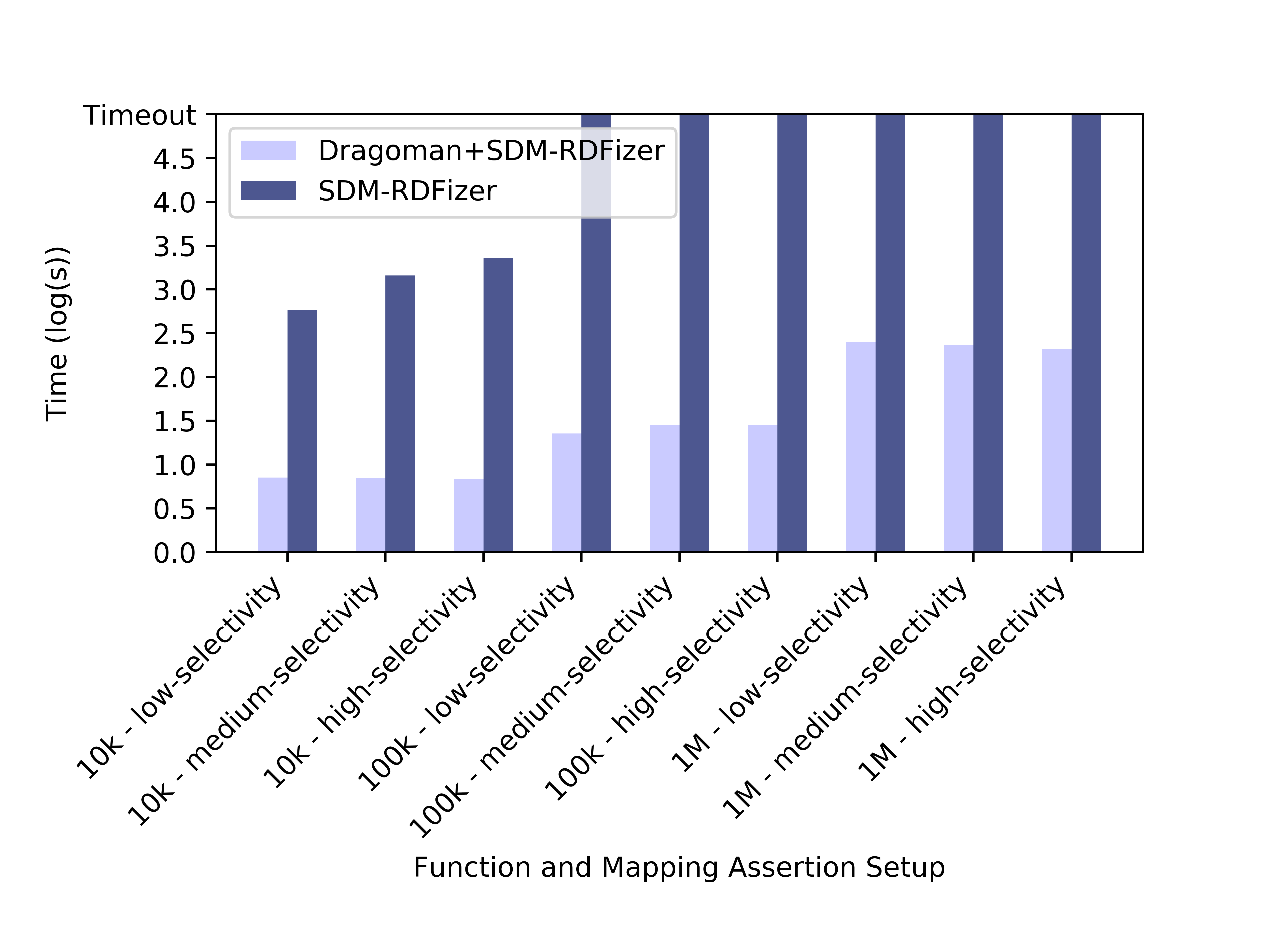}
        \label{fig:starjoin_original_simple}
    }
    \subfloat[Star Join - Composite Function ]{
        \includegraphics[trim=0 0 0 21,clip,width=0.5\linewidth]{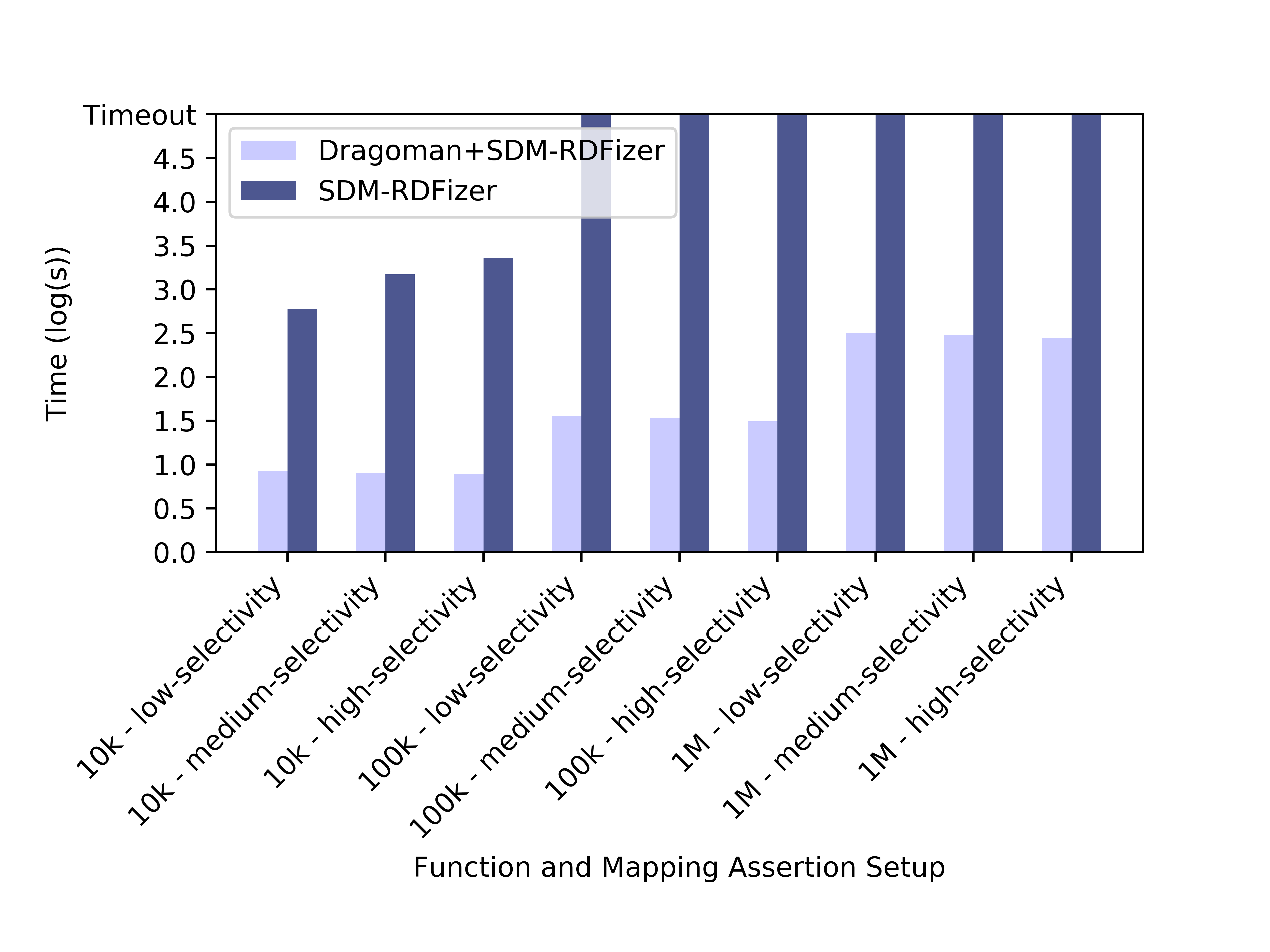}
        \label{fig:starjoin_original_composite}
    }
    \caption{\textbf{}}
    \label{fig:complexmappings}
\end{figure*}

\noindent In overall, we set up nine knowledge graph creation pipelines composed of join stars; we aim to observe the cost of executing functions using Dragoman in such expensive pipelines. In these testbeds three different values of Par2, i.e., low, medium, and high join selectivity are considered. Note that in these nine testbeds, functions are not repeated. Furthermore, in this group of experiments, we also study Par4 by including the user-defined functions in the concept mapping assertion of the first nine experiments, contrary to the previous experiments in which the same functions appear as role mapping assertions.\\ 

\noindent\textbf{Observations and Result.}
\autoref{fig:chainJoin} summarizes the results of the experiments in the chain join category, while, \autoref{fig:starjoin_original_simple} illustrates the results of the experiments in the star join category. Considering the wide range of execution time values of these experiments, we visualize the results' logarithm value for clarity.\\

\noindent \textbf{Observations on Chain Join; Par1, Par2, and Par7:} As it can be observed in \autoref{fig:chainJoin}, applying Dragoman significantly improves all knowledge graph creation pipelines' performance. Contrary to the previous group of experiments, i.e., \autoref{sec:efficiency}, these experiments, including chain join, show improvement independent of the size of the data sources. Another important observation to note is the impact of join selectivity. Although the quantity of savings in experiments with different join selectivity differs, the reduced time is considered in all the cases. \\

\noindent \textbf{Observations on Star Join, Par1, Par2, Par4 and Par6:} Similar to the results obtained from chain join experiments, as it can be observed in \autoref{fig:starjoin_original_simple}, utilizing Dragoman improves the performance of knowledge graph creation pipelines in all the nine cases which include star joins in their original mapping assertions. In other words, the only difference in the observed results of the setups with various values for Par1 or Par2 is the quantity of the savings; the higher the data size or selectivity, the greater the savings. 

\subsubsection{Composite Functions}
\label{sec:composite_functions}
\noindent \textbf{Testbeds.} In this group of experiments, we aim to study the impact of composite user-defined functions in mapping assertions, i.e., Par9. To this end, we repeat the same nine experiments of the first category of star join explained earlier. We repeat these nine experiments explained in \autoref{sec:complex_join} with composite functions in the form of $A(B(C(.)))$.\\ 

\noindent\textbf{Observation and Results.} In addition to the results of the experiments with simple functions illustrated in \autoref{fig:starjoin_original_simple}, the results of the same experiments with composite functions can be observed in \autoref{fig:starjoin_original_composite}. Comparing \autoref{fig:starjoin_original_simple} and \autoref{fig:starjoin_original_composite}, we can conclude that the complexity of a function in terms of being composite has no significant impact on the overall execution time of the knowledge graph creation pipeline.

\subsection{Discussion: Results of Experimental Study}
\noindent \textbf{RQ1.} The answer to this research question is studied in all three categories of the experiments by dividing the knowledge graph creation process into two tasks. The impact of performing the task of functions evaluation and transforming the DIS into a function-free one by Dragoman is compared with the performance of two other RML-compliant engines for the same tasks. The common observation among most of the experiments shows that in the case of having a small data source, i.e., 10k, the optimization performed by Dragoman costs more than the amount of saving that it presents to the process of knowledge graph creation. In contrast, applying Dragoman in creating knowledge graphs from larger data sources, i.e., 100k and 1M, optimizes the overall cost. The ``complex multi-sources role mapping assertions'' category of the experiments is exempted from explained conclusions; they show savings even with small datasets.  

\noindent \textbf{RQ2.} \textbf{Data size.} As expected, based on similar studies, the data source size is an important factor in deciding the application of Dragoman. It should not be interpreted that the size of the data source impacts the performance of the Dragoman, but rather it impacts the number of optimization benefits that are brought by Dragoman. \textbf{Join selectivity.} As expected, the results of the ``complex multi-sources role mapping assertions'' illustrate that join selectivity is an effective parameter in knowledge graph creation pipelines.
In contrast, the results of the same experiments which utilize Dragoman show no significant difference in the execution time between data sources with various join selectivity. \textbf{Mapping Assertions.} As expected, due to eager evaluation, Dragoman improves the execution time of knowledge graph creation processes that involve repeating the same user-defined function in different mapping assertions.
In contrast, no specific impact on the overall performance of Dragoman can be observed considering the parameter Par4, i.e., the type mapping assertion that involves user-defined functions. The effect of the parameters Par6 and Par7 are studied by the ``complex multi-sources role mapping assertions'' group of experiments. The significant improvements observed in using Dragoman in these results position Dragoman as a requirement in any knowledge graph creation pipeline, including functions and complex joins. Because even simple joins can be expensive, providing optimization for executing complex joins and transforming the DIS, which times out into an efficient one, show obvious use cases of Dragoman. \textbf{Function Parameters.} It can be observed from the results of the first category of the experiments, i.e., ``function type'', that the type of the function has no obvious impact on the overall execution time of knowledge graph creation pipelines. The result of the ``star join'' experiments reveal the same conclusion for the ``function composition'' parameter. Nonetheless, it should be noted that it also implies that the ``function composition'' parameter has no impact on the performance of Dragoman.

\section{Related Work}
\label{sec:relatedwork}
\subsection{Data Integration Systems and Knowledge Graphs}
\label{sub:dis}
\noindent Data integration is the problem of providing a unified view of the data residing in separated sources and the fundamental challenge in knowledge graph creation. Lenzerini~\cite{lenzerini2002data} formalizes the components of a DIS and represents a pivot for Ontology-Based Data Access/Integration (OBDA/I)~\cite{PoggiLCGLR08} which plays an important role in overcoming the semantic heterogeneity problem. The problem of scaling the creation of a knowledge graph as the outcome of data integration is gaining momentum~\cite{corcho2020towards}. The efforts made by the community to overcome the problem of scaled-up RDF knowledge graph creation can be grouped into two categories; \textbf{a.} resolving the data integration while generating the RDF entities and \textbf{b.} integration after modeling the data into RDF triples. \\

\noindent Starting with the category \textbf{a.}, Szekely, et al. \cite{szekely2015building} propose an approach for building knowledge graphs and devise the DIG system which resorts to KARMA \cite{knoblock2015exploiting}, a semantic DIS proposed by Knoblock et al., for integration at the level of schema. Jozashoori and Vidal define MapSDI~\cite{jozashoori2019mapsdi}, a rule-based mapping optimization for knowledge graph creation. Following the category \textbf{b.}, Collarana et al.~\cite{collarana2017minte} introduce MINTE, an integration framework that relies on the concept of RDF molecules to represent RDF entities semantically and be able to create, identify, and merge semantically equivalent RDF entities. Collarana et al. also divide available approaches into two categories; the first category includes data integration approaches focusing on linking tasks, e.g., Silk~\cite{isele2010silk}. Silk can discover links between RDF resources based on the similarity between datatype properties considering link specifications provided by users. However, the approaches in this category consider the data preparation and operation as separate steps before the main data integration process, a.k.a. pre-processing. Pre-processing steps are usually developed as \textit{ad-hoc} programs for each specific knowledge graph pipeline which fails to be traceable and maintainable in scale.\\

\noindent The second category consists of approaches that mainly contribute to the fusion task. For instance, Sieve~\cite{mendes2012sieve} proposed by Mendes et al. is a framework for quality assessment and fusion methods. LDIF is introduced by Schultz et al.~\cite{schultz2011ldif}, which relies on a set
of tools, including Silk~\cite{isele2010silk} and Sieve~\cite{mendes2012sieve} to
link identified entities and the data fusion
tasks, respectively. Benbernou et al.~\cite{benbernou2017semantic} propose a semantic-based RDF data fusion relying on an inference mechanism using rules. Nevertheless, the approaches explained in this category consider the data operations, such as entity alignment, as a differentiated step after the main integration process, a.k.a. post-processing. Hence, these approaches ignore the cost forced on the knowledge graph creation process due to generating the same nodes multiple times. These limitations in both categories shed light on the necessity of integrating the data operation in the main data integration process and knowledge graph creation. Consequently, it is essential to introduce scaled-up approaches to build knowledge graphs from the data integration systems that involve data operation functions.

\subsection{Mapping Languages}
\noindent To specify the mapping between the unified schema and the data sources transparently, a declarative mapping language can be applied. R2RML~\cite{das2012r2rml} recommended by the World Wide Web Consortium (W3C) and RDF Mapping Languages (RML)~\cite{dimou2014rml}, the extension of R2RML, are two exemplar mapping languages. To involve the data pre-processing in mappings as part of the declarative process, De Meester et al. introduce Function Ontology (FnO)~\cite{de2016ontology}. Moreover, RML provides an extension, FNML a.k.a RML+FnO, to define functions in mappings utilizing FnO definition. Defining the pre-processing functions declaratively as part of the data semantification process ensures the transparency of a knowledge graph creation pipeline. Additionally, it enhances the maintainability, reusability, and reproducibility of the data integration and transformation pipeline. Mentioned features lead to a knowledge graph creation pipeline accommodating FAIR guiding principles~\cite{wilkinson2016fair}, which makes them a good replacement for data pre-processing steps. 

\subsection{Data Integration System Executing Engines}
\noindent The semantic web community has contributed to proposing several methods and tools to translate R2RML and RML mappings and transform data into RDF model. In addition to the approaches explained in \autoref{sub:dis}, R2RML and RML engines are tackling the problem of scaled-up knowledge graph creation. For instance, Arenas-Guerrero et al. propose Morph-KGC~\cite{arenas2022morph}, an approach to partition RML rules and execute them in parallel. Iglesias et al.~\cite{iglesias2022scaling} introduce an engine-agnostic optimizing approach that improves the performance of available [R2]RML-compliant engines by planning the execution of mapping partitions. In contrast, engines able to translate RML+FnO have gained less attention and contribution. SDM-RDFizer~\cite{iglesias2020sdm}, RMLMapper~\cite{dimou_ldow_2016}, RocketRML~\cite{rocketrml}, and CARML\furl{https://github.com/carml/carml} are the only accepted examples of the engines able to translate RML+FnO. Although valuable, these engines introduce no particular optimization for function execution; they follow a lazy evaluation strategy executing FnO functions. To our knowledge, FunMap~\cite{jozashoori2020funmap} is the only engine developed specifically to evaluate functions in RML+FnO mappings and provide a function-free data integration system. However, FunMap does not address the execution of composite functions, generating different term types from the output of functions, and making decisions about the optimization plan based on the types of the provided mapping assertions. Furthermore, the extension and application of all mentioned engines, including FunMap, for newly introduced data operation functions can be complicated and discouraging. For instance, one of the unpromising requirements for users to extend the available engines with their data operation functions is to get familiar with the already existing implementation of the engines; users need to understand where and how the implementation of new functions needs to be added. All the mentioned drawbacks of available tools lead us to introduce an optimized translator of RML + Fno able to interpret and evaluate complex mappings and functions while ensuring the facilitated application and adaptation.

\subsection{Benchmarking and Studies Reporting Parameters impacting Knowledge Graph Creation Performance}
\noindent Namici et al.~\cite{namici2018comparing} compare two state-of-the-art engines in Ontology-Based Data Access by formalizing the two systems considering W3C-compliant settings. In addition to the theoretical contributions, there have been a few empirical evaluations reporting the parameters that impact a materialized knowledge graph creation pipeline, such as the study by Chaves et al.~\cite{chaves2019parameters}. Despite the importance of considering reported variables, they are not fully adopted in the available benchmarks such as GTFS-Madrid-Bench~\cite{chaves2020gtfs}, a recently introduced benchmark to evaluate the knowledge graph creation approaches. For instance, GTFS-Madrid-Bench lacks the required testbeds to study the impact of parameters such as the ``join selectivity''~\cite{chaves2019parameters}. This forces us to create a new testbed, ensuring the evaluation of missing parameters to perform a comprehensive empirical evaluation of our approach.

\section{Conclusion and Future Work}
\label{sec:conclusions}
\noindent In this work, the problem of efficiently creating a knowledge graph from a function-included data integration system is tackled. Dragoman, a system-agnostic engine, is proposed as a solution for optimization and function execution; Dragoman introduces a set of transformations. Relying on an eager evaluation, Dragoman materializes functions in mappings before deciding on the set of required transformations based on provided mapping assertions. Dragoman determines which transformations are needed to be performed on a given data integration system such that the knowledge graph generated by the transferred data integration system is the same as the knowledge graph generated by the original one, nonetheless, in less execution time. As observed in the empirical study, the application of Dragoman in knowledge graph creation pipelines from large data sources can reduce overall costs significantly; up to 75\% savings. It is also discovered that with complex mapping assertions such as ``star join'', utilizing Dragoman always decreases knowledge graph creation costs, albeit with small data sources.   
\noindent In the future, we aim to explore the optimization options for executing functions as part of declarative virtual knowledge graph creation pipelines.

\section{Acknowledgement}
\noindent This work has been partially supported by the EU H2020 RIA
funded project CLARIFY with grant agreement No 875160, P4-LUCAT with GA No. 53000015, and Federal Ministry for Economic Affairs and Energy of Germany in the project CoyPu (No 01MK21007[A-L]). Furthermore, Maria-Esther Vidal is partially supported by Leibniz Association in the program "Leibniz Best Minds: Programme for Women Professors", project TrustKG-Transforming Data in Trustable Insights with grant P99/2020.

\bibliographystyle{abbrv}
\bibliography{bibliography}

\end{document}